\documentclass[aps, prc, 12pt, onecolumn, amsmath, amssymb, superscriptaddress, nofootinbib]{revtex4-2}
\usepackage{graphicx}
\usepackage{color}
\usepackage{lineno}

\newcommand{\be}{\begin{equation}}
\newcommand{\ee}{\end{equation}}
\newcommand{\beq}{\begin{eqnarray}}
\newcommand{\eeq}{\end{eqnarray}}
\newcommand{\ba}{\begin{array}}
\newcommand{\ea}{\end{array}}

\begin{document}

\title{Universal Mass Equation for \\ Equal-Quantum Excited-States Sets~I}

\vspace{3mm}
\date{\today}

\author{\mbox{L.~David~Roper}} 
\altaffiliation{Author: \texttt{roperld@vt.edu}} 
\affiliation{Prof. Emeritus of Physics, Virginia Polytechnic Institute and State University\\
    1001 Auburn Dr. SW, Blacksburg VA 24060, USA}

\author{\mbox{Igor~Strakovsky}} 
\altaffiliation{Author: \texttt{igor@gwu.edu}} 
\affiliation{Institute for Nuclear Studies, Department of Physics, 
    The George Washington University, Washington, DC 20052, USA}

\begin{abstract}
The masses of fifteen baryon sets and twenty-four meson sets of three or more equal-quantum excited states, using Breit-Wigner PDG masses and their uncertainties at fixed $J^P$ for baryons and $J^{PC}$ for mesons, are fitted by a simple two-parameter logarithmic function, $M_n = \alpha Ln(n) + \beta$, where $n$ is the level of radial excitation. The conjecture is made that accurately measured masses of all equal-quantum baryons (including LHCb exotic $P_{c\bar{c}}^+$s) and meson excited states (including $s\bar{s}$, $s\bar{c}$, $c\bar{c}$, $c\bar{b}$, and $b\bar{b}$ states) are related by the logarithmic function used here; at least for the mass range of currently known excited states.
The baryon ``star'' rating case is evaluated.
The Cornell potential is an example of how a logarithmic behavior can be explained by an appropriate potential.
Thus, a ``universal mass equation'' (UME) for equal-quantum excited-state sets is presented.
\end{abstract}

\maketitle

\section{Introduction} 
\label{Sec:Intro}

This research started as a study of the excited states of the proton/neutron. The interesting result of that study led to studies of other equal-quantum combinations of particle-physics resonances. (The terms ``excited state'' and ``resonance'' are synonymous.) An excited state is a state in an equal-quantum excited-states set that has a higher energy than the ground state. In this work, the authors often label a ground state as an excited state of the vacuum.

Data for baryon and meson excited states are reported in the Particle Data Listings from the Particle Data Group (PDG) report ~\cite{ParticleDataGroup:2024cfk}. We use a Naming Scheme for Hadrons following Ref.~\cite{Amsler:2024}.
Resonance mass and width, with their uncertainties, are reported as Breit-Wigner (BW) values or real- and imaginary-pole positions (PP) in the complex-energy plane. The authors prefer the latter by the belief that they are closer to physical reality; however, BW mass values are reported more often than PP real parts. Therefore, BW masses are used in this document. (The BW mass is the real part of the PP on the second Riemann sheet.)

Excited-state data sufficiently accurate to do this study are available for forty-one baryon and meson data sets of three or more equal-quantum 
excited states. The uncertainties for missed and predicted states are given without correlation contributions.

Initially, the authors were reluctant to use excited-state data labeled ``OMITTED FROM SUMMARY TABLE'' by PDG; however, it was noticed that many of these data lie on or near the logarithmic curve of the data set, so such data were used.

\section{Simple Mathematics}

In 1964 the first excited state of the proton/neutron was reported by one (LDR) of the author's Ph.D. work at MIT and Lawrence Radiation Laboratory, the $1440~\mathrm{MeV}$~$P_{11}$ pion-nucleon scattering lowest mass resonance $N1/2^+$ (``Roper resonance'')~\cite{Roper:1964zza}.
Since then, four higher-mass proton/neutron excited states have been discovered ($1710~\mathrm{MeV}$, $1880~\mathrm{MeV}$, $2100~\mathrm{MeV}$, and $2300~\mathrm{MeV}$).

Since the relation between the scattering amplitudes of isospin-1/2 $\equiv A(1/2)$, in terms of $\pi^+ p \equiv A(+)$ and $\pi^- p \equiv A(-)$, is 
\begin{equation}
   A(1/2)=[3A(-)-A(+)]/2, 
\label{eq:eq1}
\end{equation}
the neutron mass ($939.565~\mathrm{MeV}$) is used as the ground-state mass in this data set instead of the proton mass ($938.272~\mathrm{MeV}$)~\cite{ParticleDataGroup:2024cfk}.

In plotting the mass data for the $N1/2^+$ data set, we found an excellent fit at fixed $J^P$ for baryons (see Sec.~\ref{Sec:BAR}) and fixed $J^{PC}$ for mesons (see Sec.~\ref{Sec:MES}), using a $\chi^2$ minimization procedure for the logarithm function
\begin{equation}
    M_n = \alpha~Ln(n) + \beta \>,
\label{eq:eq2}
\end{equation}
and used the fit to predict the masses of one missing excited state and four higher excited states of the proton/neutron. Here, $n$ is the radial excitation level and $\alpha$ and $\beta$ are free parameters. (The parameter $\beta$ is essentially the lowest mass in the data set [$\beta = M_1$, since $Ln(1) = 0$], especially if the lowest mass is the mass measured most accurately of the set, as the neutron is in the data set $N1/2^+$ (see Table~\ref{tbl:taba1}.)

The PDG provides the data mass uncertainties used in the $\chi^2$ minimization procedure, which contains a method to determine uncertainties in $\alpha$ and $\beta$ , $\delta \alpha$ and $\delta \beta$, which are used to calculate uncertainties in predicted masses using $\delta M_n=\sqrt{[Ln(n)\delta \alpha]^2+(\delta \beta)^2}$.

Note that the starting mass of a state set does not have to be the lowest mass, which may not be known. The value of parameter $\alpha$ will be smaller and the value of parameter $\beta$ will be greater if the starting state of a series is above the ground state. 

The mass difference between two nearby masses in an excited-states data set, $\Delta M_{2,1} = \alpha [Ln(n_2) - Ln(n_1)]$, which decreases as $n$ increases, approaches $0$ as $n$ approaches $\infty$. In experiments, as the mass differences get smaller, experimental uncertainties may make it impossible to measure the higher masses of an excited-data set.

\textbf{
Why the Logarithm Function?}\\
One can rewrite Eq.~(\ref{eq:eq2}) by using
\begin{equation}
    \frac{dM_n}{dn} = \frac{\alpha}{n} \>.
\label{eq:eq3}
\end{equation}
Then,
\begin{equation}
    M_n = \alpha \int \frac{1}{n} dn + \beta  \cong \alpha \sum _{n=1}^{N} \frac{\delta n}{n} + \beta \text {, for large $N$} \>.
\label{eq:eq4}
\end{equation}
So, $M_n$ can be approximated as a sum over a large number of $\frac{1}{n}$ terms, such as in energy levels. Perhaps this is the reason why the logarithm function works so well in these mass fits.

The logarithm function works very well for fitting the masses for the excited states of many equal-quantum baryon and meson data sets, as shown in the following graphs (blue circles). The radial excitation level $n$ is
unknown for each state, and we make judgments about ``missed'' states~\cite{Koniuk:1979vw} in some cases, whose masses we calculate (green triangles) and predict higher states (magenta diamonds). Extrapolation is sensitive, and for that reason, we calculate just four predicted higher masses; for some sets, higher mass states may be accurately predicted; for example, the upsilon meson ($\Upsilon$) set.

The best equal-quantum excited-states sets for observing the validity of the universal mass equation are $N1/2^+$-baryons, $\Sigma3/2^+$-baryons, $f_0$-mesons, $a_0$-mesons, and $\Upsilon$-mesons sets.

Research about the Standard Model physics of the logarithmic fits to particle-physics masses reported in this document would be welcomed by the authors. We have not found any such theory in the literature.

\section{Baryon Excited-States}
\label{Sec:BAR}

Baryon-excited-state sets analyzed: Name $J^P$~[\# of excited-states data in a set by PDG].\\
Baryons~({15} Data Sets): 
$N1/2^+$~[6], $N1/2^{-}$~[3], $N3/2^+$~[3], $N3/2^-$~[4], $N5/2^+$~[3], $N5/2^-$~[3], $\Delta 1/2^-$~[3], $\Delta 3/2^+$~[3], $\Lambda 1/2^+$~[4], $\Lambda 1/2^-$~[3], 
$\Sigma 1/2^+$~[3], $\Sigma 1/2^-$~[4], $\Sigma 3/2^+$~[5], $\Sigma 3/2^-$~[3], 
and $P_{c\bar{c}}^+$~[4].

\subsection{$N$ Baryons}

\subsubsection{$N1/2^+$ Excited-States}

Quantum numbers for these states: $I(J^P)S = 1/2(1/2^+)0$, where $I$ is isospin, $J$ is spin, $+$ or $-$ is the parity sign, and $S$ is the strangeness number.

\begin{figure}[htb!]
\centering
{
    \includegraphics[width=0.5\textwidth,keepaspectratio]{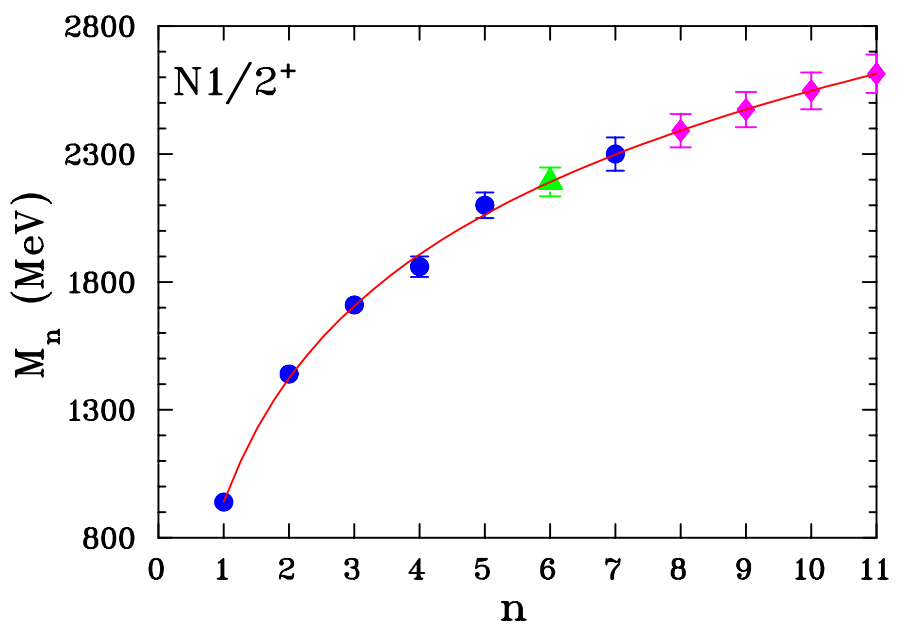} 
}

\centerline{\parbox{0.8\textwidth}{
\caption[] {\protect\small
Data for $N1/2^+$ (blue circles): $N(940)$, $N(1440)$, $N(1710)$, $N(1860)$, $N(2100)$, and $N(2300)$~\cite{ParticleDataGroup:2024cfk}.
The green triangle is the calculated mass of the missing $N(2191)$ state with mass of $2191\pm 56~\mathrm{MeV}$.
Predicted states (magenta diamonds): $N(2391)$, $N(2474)$, $N(2547)$, and $N(2614)$
with masses of $2391\pm 65~\mathrm{MeV}$, $2474\pm 69~\mathrm{MeV}$, $2547\pm 72~\mathrm{MeV}$, and $2614\pm 75~\mathrm{MeV}$, respectively.
The solid red curve presents the best-fit.
The fit parameters are $\alpha = 698.2\pm  
14.1~\mathrm{MeV}$ and $\beta = 939.6\pm 5.4\times 10^{-7}~\mathrm{MeV}$.
$\chi^2/\mathrm{DoF} = 0.6$ and CL = 68.7\%. (DoF is degree of freedom and CL is Confidence Level.) 
}
\label{fig:fig1} } }
\end{figure}

The logarithmic fit to the neutron mass and BW masses (MeV) of the six known excited states $N1/2^+$ (blue circles) and four projected higher excited states (magenta diamonds) is shown in Fig.~\ref{fig:fig1}.  In addition, a missing state (green triangle) is shown as calculated.

The four projected higher-mass proton/neutron excited states may be difficult to measure in pion-nucleon scattering experiments, especially if they are masked by other nearby resonances, similar to the case of the $N(1440)$ (Roper) resonance.
Of course, the mass equation, Eq.~(\ref{eq:eq2}), could be used to predict even higher excited mass states for the $N1/2^+$ data set. \\

\subsubsection{$N1/2^-$ Excited-States}

A series of $N1/2^-$ pion-nucleon-scattering resonances is recorded in the Particle Data
Listings~\cite{ParticleDataGroup:2024cfk}. $N1/2^-$: $I(J^P)S = 1/2(1/2^-)0$.

\begin{figure}[htb!]
\centering
{
    \includegraphics[width=0.5\textwidth,keepaspectratio]{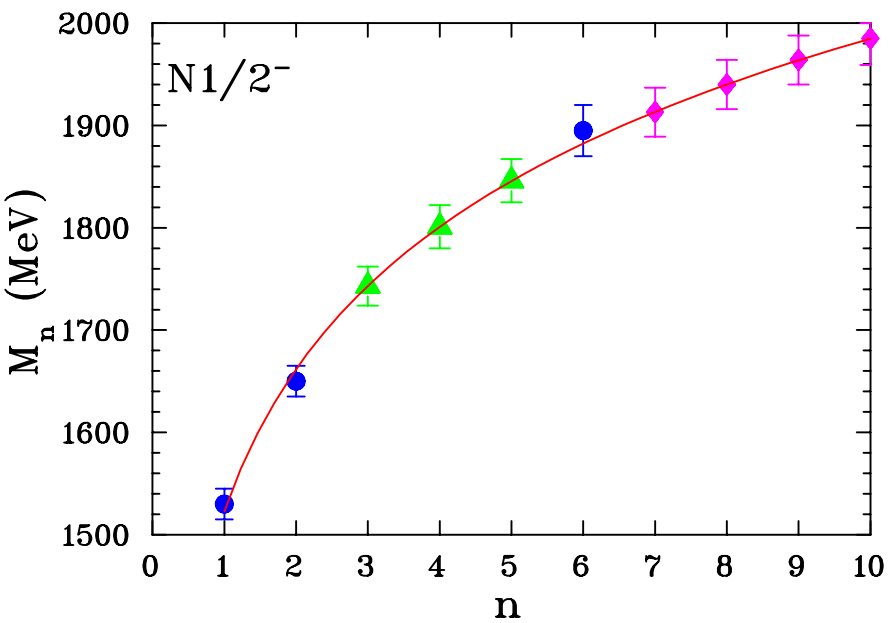} 
}

\centerline{\parbox{0.8\textwidth}{
\caption[] {\protect\small
Data for $N1/2^-$ (blue circles): $N(1530)$, $N(1650)$, and $N(1895)$~\cite{ParticleDataGroup:2024cfk}.
The green triangles are the calculated masses of the missing $N(1743)$, $N(1801)$, and $N(1846)$ states with masses of $1743\pm 19~\mathrm{MeV}$, $1801\pm 21~\mathrm{MeV}$, and $1846\pm 21~\mathrm{MeV}$, respectively.
Predicted states (magenta diamonds): $N(1913)$, $N(1940)$, $N(1964)$, and $N(1985)$
with masses of $1913\pm 24~\mathrm{MeV}$, $1940\pm 24~\mathrm{MeV}$, $1964\pm 24~\mathrm{MeV}$, and $1985\pm 26~\mathrm{MeV}$, respectively.
The solid red curve presents the best-fit.
The fit parameters are $\alpha = 200.7\pm 16.0~\mathrm{MeV}$ and $\beta = 1522.7\pm 
13.3~\mathrm{MeV}$.
$\chi^2/\mathrm{DoF} = 1.1$ and CL = 29.1\%.
}
\label{fig:fig2} } }
\end{figure}

The logarithmic fit to the BW masses (MeV) of the three known excited states
of $N1/2^-$ (blue circles) and four higher excited states (magenta diamonds) 
projected is shown in Fig.~\ref{fig:fig2}.  In addition, three missed states
(green triangles) are shown as calculated.

\subsubsection{$N3/2^+$ Excited-States}

A series of $N3/2^+$ pion-nucleon-scattering resonances is recorded in the Particle Data
Listings~\cite{ParticleDataGroup:2024cfk}. $N3/2^+$: $I(J^P)S = 1/2(3/2^+)0$.

\begin{figure}[htb!]
\centering
{
    \includegraphics[width=0.5\textwidth,keepaspectratio]{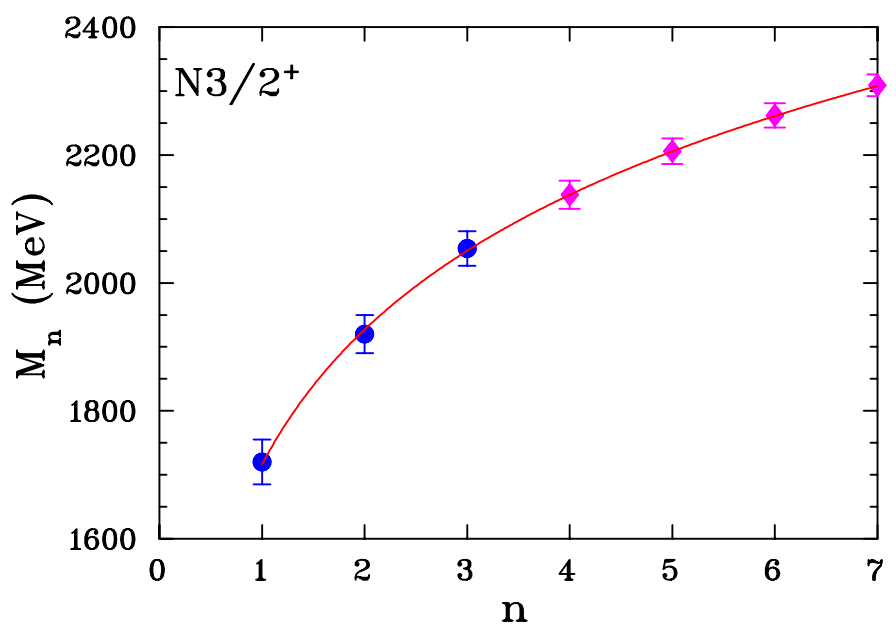} 
}

\centerline{\parbox{0.8\textwidth}{
\caption[] {\protect\small
Data for $N3/2^+$ (blue circles): $N(1720)$, $N(1920)$, and $N(2054)$~\cite{ParticleDataGroup:2024cfk}.
Predicted states (magenta diamonds): $N(2138)$, $N(2206)$, $N(2262)$, and $N(2309)$ with masses of $2138\pm 22~\mathrm{MeV}$, $2206\pm 20~\mathrm{MeV}$, $2262\pm 19~\mathrm{MeV}$, and $2309\pm 17~\mathrm{MeV}$, respectively.
The solid red curve presents the best-fit.
The fit parameters are $\alpha = 304.4\pm 39.1~\mathrm{MeV}$ and $\beta = 1716.3\pm 
32.7~\mathrm{MeV}$.
$\chi^2/\mathrm{DoF} = 0.1$ and CL = 76.7\%.
}
\label{fig:fig3} } }
\end{figure}

The logarithmic fit to the BW masses (MeV) of the three known excited states $N3/2^+$ (blue circles) and the four higher excited states (magenta diamonds) projected is shown in Fig.~\ref{fig:fig3}.

\subsubsection{$N3/2^-$ Excited-States}

A series of $N3/2^-$ pion-nucleon-scattering resonances is recorded in the Particle Data
Listings~\cite{ParticleDataGroup:2024cfk}. $N3/2^-$: $I(J^P)S = 1/2(3/2^-)0$.

\begin{figure}[htb!]
\centering
{
    \includegraphics[width=0.5\textwidth,keepaspectratio]{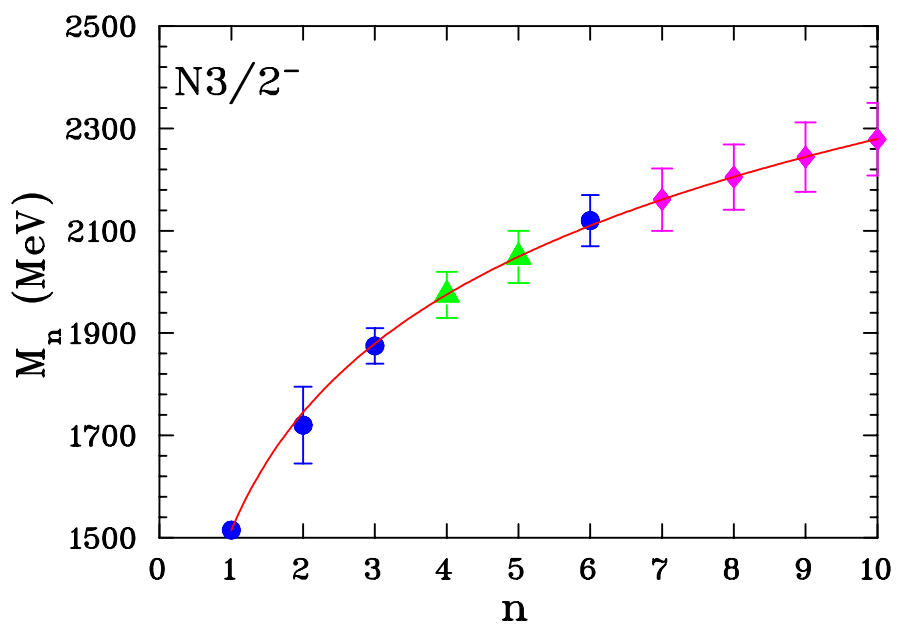} 
}

\centerline{\parbox{0.8\textwidth}{
\caption[] {\protect\small
Data for $N3/2^-$ (blue circles): $N(1515)$, $N(1720)$, $N(1875)$, and $N(2120)$~\cite{ParticleDataGroup:2024cfk}.
The green triangles are the calculated masses of the missing $N(1975)$ and $N(2049)$ states
with masses of $1975\pm 45~\mathrm{MeV}$ and $2049\pm 51~\mathrm{MeV}$, respectively.
Predicted states (magenta diamonds): $N(2161)$, $N(2205)$, $N(2244)$, and $N(2279)$ with masses of $2161\pm 61~\mathrm{MeV}$, $2205\pm 64~\mathrm{MeV}$, $2244\pm 68~\mathrm{MeV}$, and $2279\pm 71~\mathrm{MeV}$, respectively.
The solid red curve presents the best-fit.
The fit parameters are $\alpha = 332.0\pm 20.9~\mathrm{MeV}$ and $\beta = 1514.9\pm 
5.0~\mathrm{MeV}$.
$\chi^2/\mathrm{DoF} = 0.1$ and CL = 91.8\%.
}
\label{fig:fig4} } }
\end{figure}

The logarithmic fit to the BW masses (MeV) of the four known excited states of $N3/2^-$ (blue circles) and four higher excited states (magenta diamonds) projected is shown in Fig.~\ref{fig:fig4}.  In addition, two missed states (green triangles) are shown as calculated.

\subsubsection{$N5/2^+$ Excited-States}

A series of $N5/2^+$ pion-nucleon scattering resonances is recorded in the Particle Data Listings~\cite{ParticleDataGroup:2024cfk}. $N5/2^+$: $I(J^P)S = 1/2(5/2^+)0$. 

\begin{figure}[htb!]
\centering
{
    \includegraphics[width=0.5\textwidth,keepaspectratio]{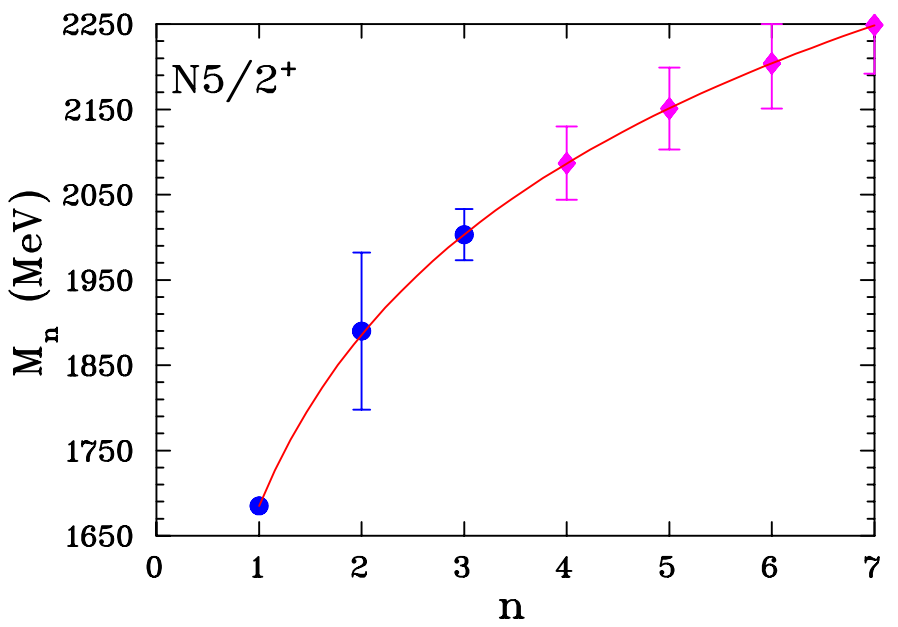} 
}

\centerline{\parbox{0.8\textwidth}{
\caption[] {\protect\small
Data for $N5/2^+$ (blue circles): $N(1685)$, $N(1890)$, and $N(2003)$~\cite{ParticleDataGroup:2024cfk}.
Predicted states (magenta diamonds): $N(2087)$, $N(2151)$, $N(2204)$, and $N(2249)$ with masses of $2087\pm 43~\mathrm{MeV}$, $2151\pm 48~\mathrm{MeV}$, $2204\pm 53~\mathrm{MeV}$, and $2249\pm 57~\mathrm{MeV}$, respectively.
The solid red curve presents the best-fit.
The fit parameters are $\alpha = 287.7\pm 27.1~\mathrm{MeV}$ and $\beta = 1685.0\pm 
5.0~\mathrm{MeV}$.
$\chi^2/\mathrm{DoF} = 0.002$ and CL = 96.3\%.
}
\label{fig:fig5} } }
\end{figure}

The logarithmic fit to the BW masses (MeV) of the three known excited states $N5/2^+$ (blue circles) and the four higher excited states projected (magenta diamonds) is shown in Fig.~\ref{fig:fig5}.

\subsubsection{$N5/2^-$ Excited-States}

A series of $N5/2^-$ pion-nucleon scattering resonances is recorded in the Particle Data 
Listings~\cite{ParticleDataGroup:2024cfk}. $N5/2^-$: $I(J^P)S = 1/2(5/2^-)0$. 

\begin{figure}[htb!]
\centering
{
    \includegraphics[width=0.5\textwidth,keepaspectratio]{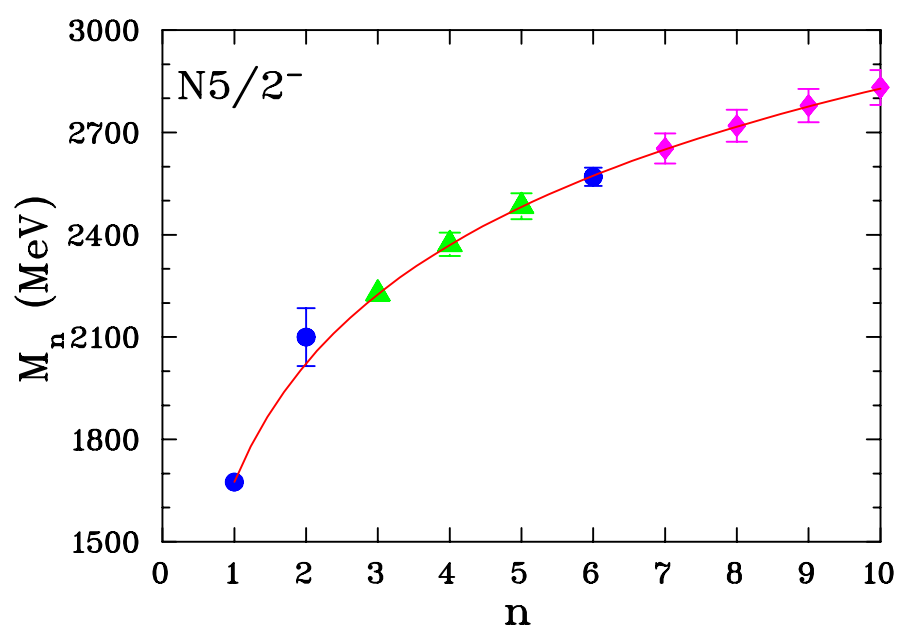} 
}

\centerline{\parbox{0.8\textwidth}{
\caption[] {\protect\small
Data for $N5/2^-$ (blue circle circles): $N(1675)$, $N(2100)$, and $N(2570)$~\cite{ParticleDataGroup:2024cfk}.
The green triangles are the calculated masses of the missing $N(2226)$, $N(2370)$, and 
$N(2482)$ states with masses of $2226\pm 21~\mathrm{MeV}$, $2370\pm 25~\mathrm{MeV}$, and $2482\pm 27~\mathrm{MeV}$, respectively.
Predicted states (magenta diamonds): $N(2650)$, $N(2717)$, $N(2776)$, and $N(2829)$ with masses of $2650\pm 31~\mathrm{MeV}$, $2717\pm 33~\mathrm{MeV}$, $2776\pm 34~\mathrm{MeV}$, and $2829\pm 36~\mathrm{MeV}$, respectively.
The solid red curve presents the best-fit.
The fit parameters are $\alpha = 501.0\pm 15.6~\mathrm{MeV}$ and $\beta = 1675.4\pm 
8.0~\mathrm{MeV}$.
$\chi^2/\mathrm{DoF} = 0.8$ and CL = 35.8\%.
}
\label{fig:fig6} } }
\end{figure}

The logarithmic fit to the BW masses (MeV) of the three known excited states $N5/2^-$ (blue circles) and the four higher excited states projected (magenta diamonds) is shown in Fig.~\ref{fig:fig6}.  In addition, three missed states (green triangles) are shown as calculated.

\subsection{Delta Baryons}

\subsubsection{$\Delta 1/2^-$ Excited-States}

A series of $\Delta 1/2^-$ pion-nucleon scattering resonances is recorded in the Particle Data 
Listings~\cite{ParticleDataGroup:2024cfk}. $\Delta 1/2^-$: $I(J^P)S = 3/2(1/2^-)0$. 

\begin{figure}[htb!]
\centering
{
    \includegraphics[width=0.5\textwidth,keepaspectratio]{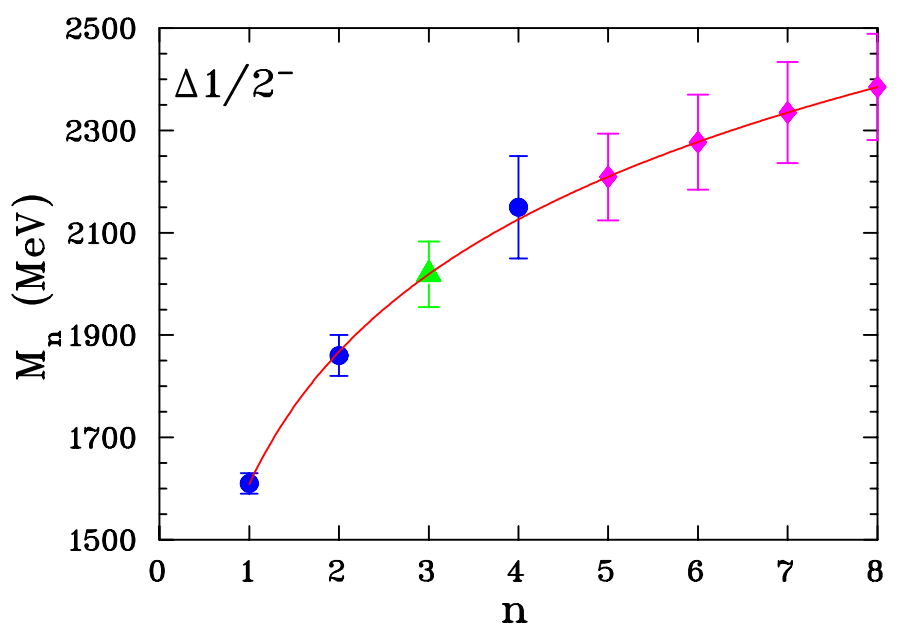} 
}

\centerline{\parbox{0.8\textwidth}{
\caption[] {\protect\small
Data for $\Delta 1/2^-$ (blue circles): $\Delta(1610)$, $\Delta(1860)$, and $\Delta(2150)$~\cite{ParticleDataGroup:2024cfk}.
The green triangle is the calculated mass of the missing $\Delta(2016)$ state with mass  of $2019\pm 64~\mathrm{MeV}$.
Predicted states (magenta diamonds): $\Delta(2209)$, $\Delta(2277)$, $\Delta(2335)$, and $\Delta(2385)$ with masses of $2209\pm 85~\mathrm{MeV}$, $2277\pm 93~\mathrm{MeV}$, $2335\pm 99~\mathrm{MeV}$, and $2385\pm 104~\mathrm{MeV}$, respectively.
The solid red curve presents the best-fit.
The fit parameters are $\alpha = 373.0\pm 50.6~\mathrm{MeV}$ and $\beta = 1609.0\pm 
19.8~\mathrm{MeV}$.
$\chi^2/\mathrm{DoF} = 0.1$ and CL = 75.8\%.
}
\label{fig:fig7} } }
\end{figure}

The logarithmic fit to the BW masses (MeV) of the three known $\Delta1/2^-$ (blue circles) excited states and one added missing state (green triangle) and four higher excited states projected (magenta diamonds) is shown in Fig.~\ref{fig:fig7}.
In addition, a missed state (green triangle) is shown as calculated.

\subsubsection{$\Delta 3/2^+$ Excited-States}

A series of $\Delta 3/2^+$ pion-nucleon scattering resonances is recorded in the Particle Data 
Listings~\cite{ParticleDataGroup:2024cfk}. $\Delta 3/2^+$: $I(J^P)S = 3/2(3/2^+)0$. 

\begin{figure}[htb!]
\centering
{
    \includegraphics[width=0.5\textwidth,keepaspectratio]{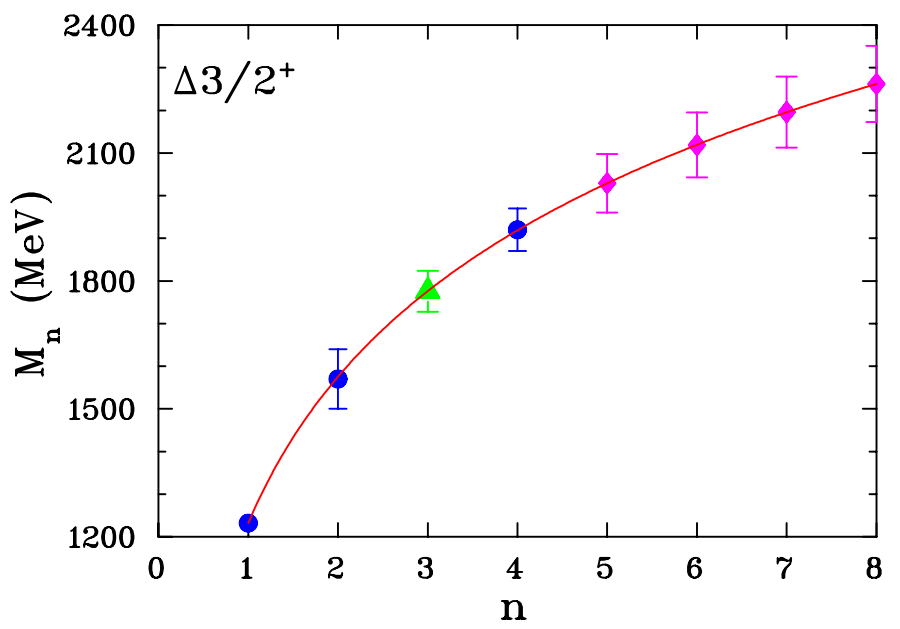} 
}

\centerline{\parbox{0.8\textwidth}{
\caption[] {\protect\small
Data for $\Delta 3/2^+$ (blue circle circles): $\Delta(1232)$, $\Delta(1570)$, and $\Delta(1920)$~\cite{ParticleDataGroup:2024cfk}.
The green triangle is the calculated mass of the missing $\Delta(1776)$ state with mass of $1776\pm 48~\mathrm{MeV}$.
Predicted states (magenta diamonds): $\Delta(2029)$, $\Delta(2119)$, $\Delta(2196)$, and $\Delta(2262)$ with masses of $2029\pm 69~\mathrm{MeV}$, $2119\pm 76~\mathrm{MeV}$, $2196\pm 83~\mathrm{MeV}$, and $2262\pm 89~\mathrm{MeV}$, respectively.
The solid red curve presents the best-fit.
The fit parameters are $\alpha = 495.3\pm 34.0~\mathrm{MeV}$ and $\beta = 1232.0\pm 
2.0~\mathrm{MeV}$.
$\chi^2/\mathrm{DoF} = 0.1$ and CL = 93.6\%.
}
\label{fig:fig8} } }
\end{figure}

The logarithmic fit to the BW masses (MeV) of the three known $\Delta 3/2^+$ (blue circles) excited states and one added missing state (green triangle) and four higher excited states projected (magenta diamonds) is shown in Fig.~\ref{fig:fig8}.
In addition, a missed state (green triangle) is shown as calculated.

\subsection{Lambda Baryons}

\subsubsection{$\Lambda$1/2$^+$ Excited-States}

A series of $\Lambda 1/2^+$ pion-nucleon scattering resonances is recorded in the Particle Data Listings~\cite{ParticleDataGroup:2024cfk}. $\Lambda 1/2^+$: $I(J^P)S = 0(1/2^+)-1$. 

\begin{figure}[htb!]
\centering
{
    \includegraphics[width=0.5\textwidth,keepaspectratio]{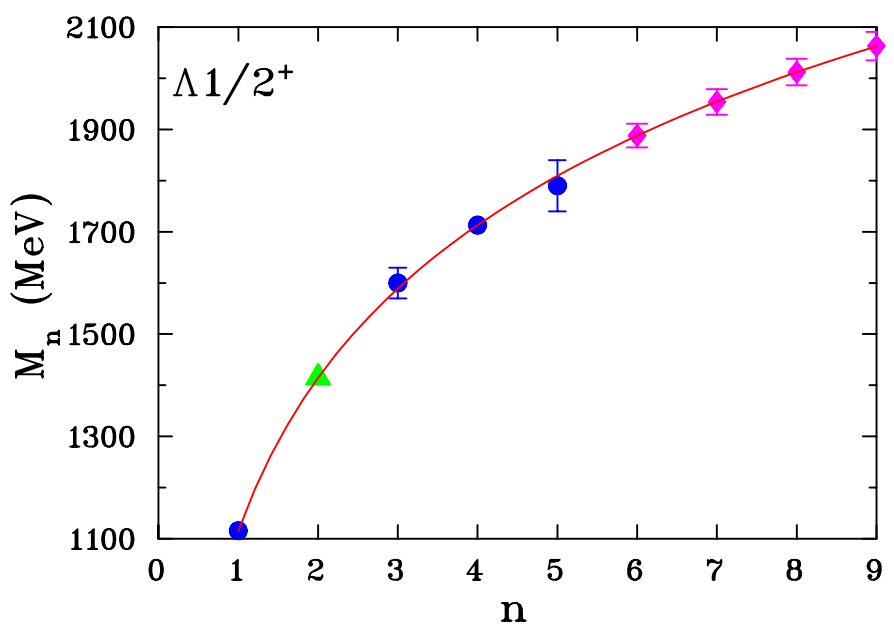} 
}

\centerline{\parbox{0.8\textwidth}{
\caption[] {\protect\small
Data for $\Lambda 1/2^+$ (blue circles): $\Lambda(1116)$, $\Lambda(1600)$, $\Lambda(1713)$, and $\Lambda$(1790)~\cite{ParticleDataGroup:2024cfk}.
The green triangle is the calculated mass of the missing $\Lambda$(1414) state with mass of $1414\pm 9~\mathrm{MeV}$.
Predicted states (magenta diamonds): $\Lambda(1888)$, $\Lambda(1954)$, $\Lambda(2012)$, and $\Lambda(2063)$ with masses of $1888\pm 23~\mathrm{MeV}$, $1954\pm 25~\mathrm{MeV}$, $2012\pm 26~\mathrm{MeV}$, and $2063\pm 28~\mathrm{MeV}$, respectively.
The solid red curve presents the best-fit.
The fit parameters are $\alpha = 430.9\pm 8.5~\mathrm{MeV}$ and $\beta = 1115.700\pm 
0.006~\mathrm{MeV}$.
$\chi^2/\mathrm{DoF} = 0.1$ and CL = 86.9\%.
}
\label{fig:fig9} } }
\end{figure}

The logarithmic fit to the BW masses (MeV) of the four known excited states $\Lambda 1/2^+$ (blue circles) and one added missing state (green triangle) and four projected higher excited states (magenta diamonds) is shown in Fig.~\ref{fig:fig9}. In addition, a missed state (green triangle) is shown as calculated.

\subsubsection{$\Lambda 1/2^-$ Excited-States}

A series of $\Lambda 1/2^-$ pion-nucleon scattering resonances is recorded in the Particle Data
Listings~\cite{ParticleDataGroup:2024cfk}. $\Lambda 1/2^-$: $I(J^P)S = 0(1/2^-)-1$. 

\begin{figure}[htb!]
\centering
{
    \includegraphics[width=0.5\textwidth,keepaspectratio]{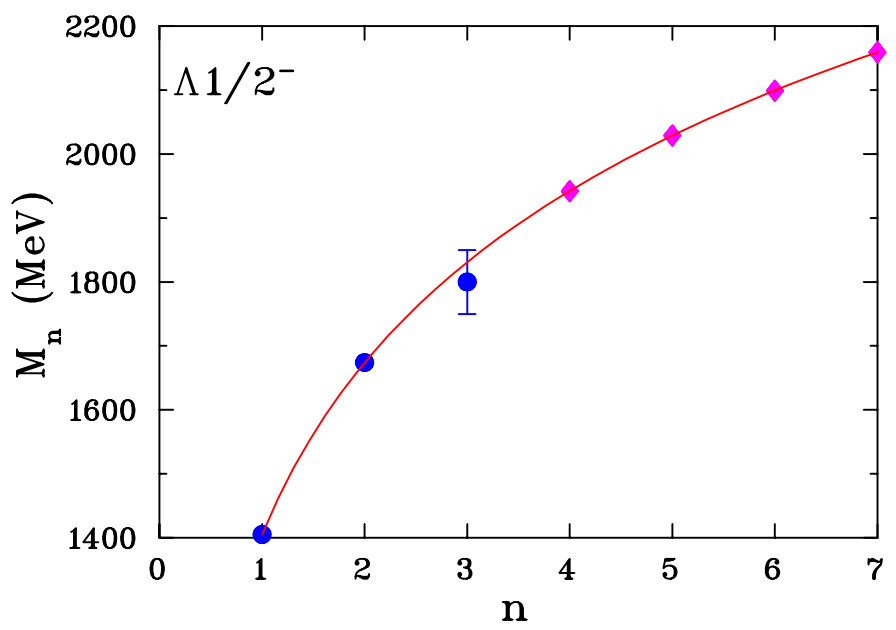} 
}

\centerline{\parbox{0.8\textwidth}{
\caption[] {\protect\small
Data for $\Lambda 1/2^-$ (blue circles): $\Lambda(1405)$, $\Lambda(1674)$, and $\Lambda$(1800)~\cite{ParticleDataGroup:2024cfk}. Predicted states (magenta diamonds): $\Lambda(1942)$, $\Lambda(2029)$, $\Lambda(2099)$, and $\Lambda(2159)$ with masses of $1942\pm 8~\mathrm{MeV}$, $2029\pm 9~\mathrm{MeV}$, $2099\pm 10~\mathrm{MeV}$, and $2159\pm 10~\mathrm{MeV}$, respectively.
The solid red curve presents the best-fit.
The fit parameters are $\alpha = 387.5\pm 6.0~\mathrm{MeV}$ and $\beta = 1405.1\pm 
1.3~\mathrm{MeV}$.
$\chi^2/\mathrm{DoF} = 0.4$ and CL = 53.5\%.
}
\label{fig:fig10} } }
\end{figure}

The logarithmic fit to the BW masses (MeV) of the three known excited states $\Lambda 1/2^-$ (blue circles) and the four higher excited states projected (magenta diamonds) is shown in Fig.~\ref{fig:fig10}.
\subsection{Sigma Baryons}

\subsubsection{$\Sigma 1/2^+$ Excited-States}

A series of $\Sigma 1/2^+$ pion-nucleon scattering resonances is recorded in the Particle Data Listings~\cite{ParticleDataGroup:2024cfk}. $\Sigma 1/2^+$: $I(J^P)S = 1(1/2^+)-1$. 

\begin{figure}[htb!]
\centering
{
    \includegraphics[width=0.5\textwidth,keepaspectratio]{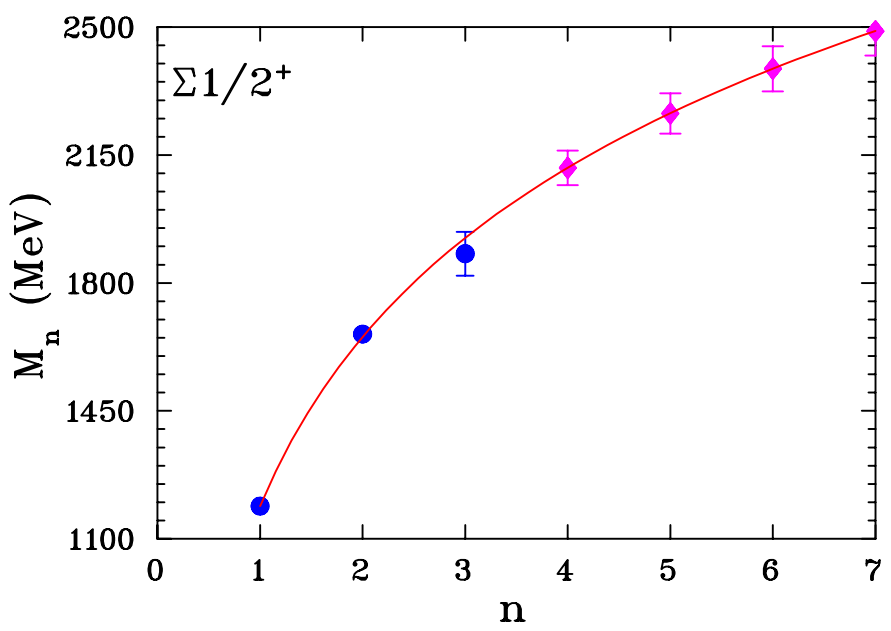} 
}

\centerline{\parbox{0.8\textwidth}{
\caption[] {\protect\small
Data for $\Sigma 1/2^+$ (blue circles): $\Sigma(1189)$, $\Sigma(1660)$, and $\Sigma(1880)$~\cite{ParticleDataGroup:2024cfk}.
Predicted states (magenta diamonds): $\Sigma(2115)$, $\Sigma(2264)$, $\Sigma(2386)$, and $\Sigma(2489)$ with masses of $2115\pm 47~\mathrm{MeV}$, $2264\pm 55~\mathrm{MeV}$, $2386\pm 62~\mathrm{MeV}$, and $2489\pm 67~\mathrm{MeV}$, respectively.
The solid red curve presents the best-fit.
The fit parameters are $\alpha = 668.0\pm 25.5~\mathrm{MeV}$ and $\beta = 1189.40\pm 
0.07~\mathrm{MeV}$.
$\chi^2/\mathrm{DoF} = 0.7$ and CL = 41.5\%.
}
\label{fig:fig12} } }
\end{figure}

The logarithmic fit to the BW masses (MeV) of the three known excited states $\Sigma 1/2^+$ (blue circles) and the four higher excited states projected (magenta diamonds) is shown in Fig.~\ref{fig:fig12}.

\subsubsection{$\Sigma 1/2^-$ Excited-States}

A series of $\Sigma 1/2^-$ pion-nucleon scattering resonances is recorded in the Particle Data 
 Listings~\cite{ParticleDataGroup:2024cfk}. $\Sigma 1/2^-$: $I(J^P)S = 1(1/2^-)-1$. 

\begin{figure}[htb!]
\centering
{
    \includegraphics[width=0.5\textwidth,keepaspectratio]{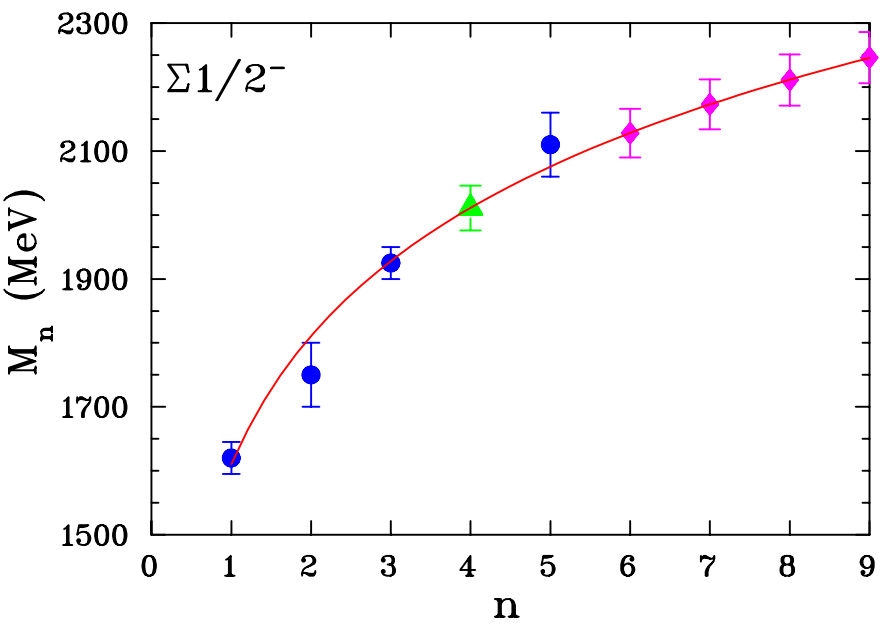} 
}

\centerline{\parbox{0.8\textwidth}{
\caption[] {\protect\small
Data for $\Sigma 1/2^-$ (blue circles): $\Sigma(1620)$, 
$\Sigma(1750)$, $\Sigma(1925)$, and $\Sigma(2110)$~\cite{ParticleDataGroup:2024cfk}.
The green triangle is the calculated mass of the missing $\Sigma$(2011) state with mass  of $2011\pm 35~\mathrm{MeV}$.
Predicted states (magenta diamonds): $\Sigma(2128)$, $\Sigma(2173)$, $\Sigma(2211)$, and $\Sigma(2246)$ with masses of $2128\pm 38~\mathrm{MeV}$, $2173\pm 39~\mathrm{MeV}$, $2211\pm 40~\mathrm{MeV}$, and $2246\pm 40~\mathrm{MeV}$, respectively.
The solid red curve presents the best-fit.
The fit parameters are $\alpha = 289.1\pm 27.1~\mathrm{MeV}$ and $\beta = 1610.4\pm 
24.0~\mathrm{MeV}$.
$\chi^2/\mathrm{DoF} = 1.1$ and CL = 34.8\%.
}
\label{fig:fig13} } }
\end{figure}

The logarithmic fit to the BW masses (MeV) of the three known excited states $\Sigma 1/2^-$ (blue circles) and the four higher excited states projected (magenta diamonds) is shown in Fig.~\ref{fig:fig13}.  In addition, a missed state (green triangle) is shown as calculated.

\subsubsection{$\Sigma 3/2^+$ Excited-States}

A series of $\Sigma 3/2^+$ pion-nucleon scattering resonances is recorded in the Particle Data 
 Listings~\cite{ParticleDataGroup:2024cfk}. $\Sigma 3/2^+$: $I(J^P)S = 1(3/2^+)-1$. 

\begin{figure}[htb!]
\centering
{
    \includegraphics[width=0.5\textwidth,keepaspectratio]{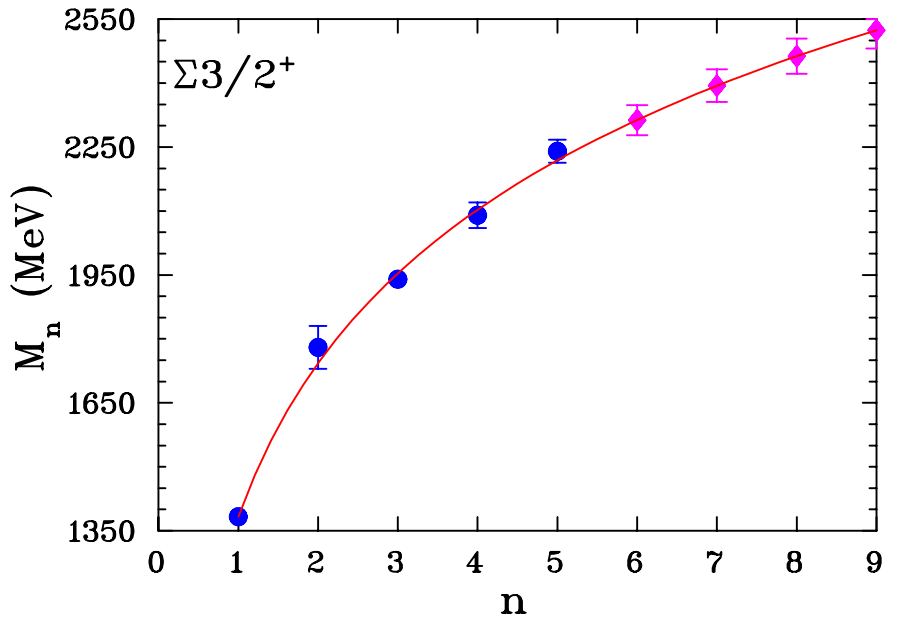} 
}

\centerline{\parbox{0.8\textwidth}{
\caption[] {\protect\small
Data for $\Sigma 3/2^+$ (blue circles): $\Sigma(1383)$, $\Sigma(1780)$, $\Sigma(1940)$, $\Sigma(2090)$, and $\Sigma$(2240)~\cite{ParticleDataGroup:2024cfk}.
Predicted states (magenta diamonds): $\Sigma(2313)$, $\Sigma(2394)$, $\Sigma(2463)$, and $\Sigma(2524)$ with masses of $2313\pm 35~\mathrm{MeV}$, $2394\pm 38~\mathrm{MeV}$, $2463\pm 41~\mathrm{MeV}$, and $2524\pm 43~\mathrm{MeV}$, respectively.
The solid red curve presents the best-fit.
The fit parameters are $\alpha = 519.4\pm 10.6~\mathrm{MeV}$ and $\beta = 1382.8\pm 0.3~\mathrm{MeV}$.
$\chi^2/\mathrm{DoF} = 0.6$ and CL = 61.4\%.
}
\label{fig:fig14} } }
\end{figure}

The logarithmic fit to the BW masses (MeV) of the five known $\Sigma 3/2^+$ excited states (blue circles) and four higher excited states projected (magenta diamonds) is shown in Fig.~\ref{fig:fig14}.

\subsubsection{$\Sigma 3/2^-$ Excited-States}

A series of $\Sigma 3/2^-$ pion-nucleon scattering resonances is recorded in the Particle Data 
 Listings~\cite{ParticleDataGroup:2024cfk}. $\Sigma 3/2^-$: $I(J^P)S = 1(3/2^-)-1$. 

\begin{figure}[htb!]
\centering
{
    \includegraphics[width=0.5\textwidth,keepaspectratio]{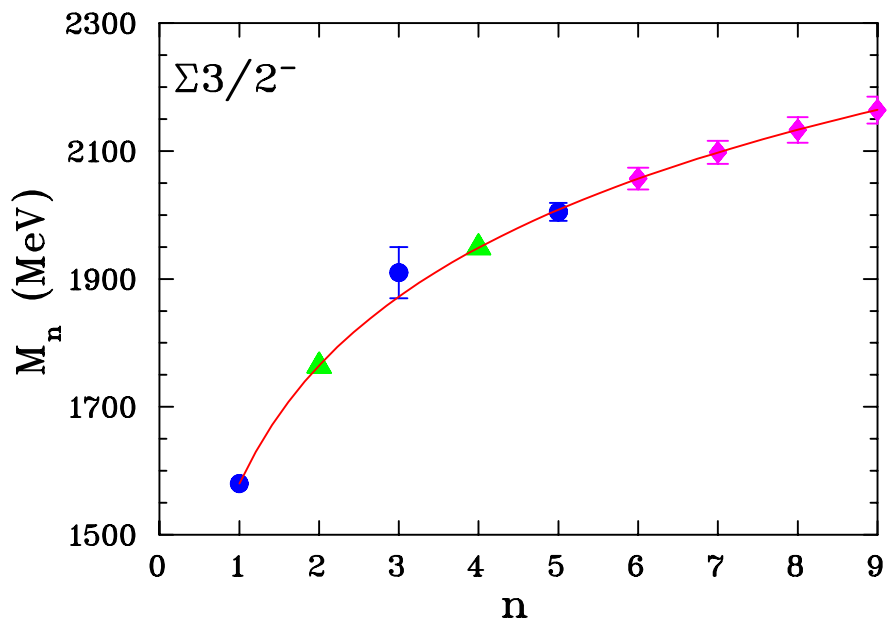} 
}

\centerline{\parbox{0.8\textwidth}{
\caption[] {\protect\small
Data for $\Sigma 3/2^-$ (blue circles): $\Sigma(1580)$, $\Sigma(1910)$, and $\Sigma(2005)$~\cite{ParticleDataGroup:2024cfk}.
A state reported by PDG are out of the sequence $\Sigma(1675)$.
The green triangles are the calculated masses of the missing $\Sigma(1764)$ and $\Sigma(1949)$ states with masses of $1764\pm 10~\mathrm{MeV}$ and $1949\pm 15~\mathrm{MeV}$, respectively.
Predicted states (magenta diamonds): $\Sigma(2057)$, $\Sigma(2098)$, $\Sigma(2133)$, and $\Sigma(2164)$ with masses of $2057\pm 17~\mathrm{MeV}$, $2098\pm 18~\mathrm{MeV}$, $2133\pm 20~\mathrm{MeV}$, and $2164\pm 21~\mathrm{MeV}$, respectively.
The solid red curve presents the best-fit.
The fit parameters are $\alpha = 265.9\pm 8.8~\mathrm{MeV}$ and $\beta = 1580.1\pm 
4.0~\mathrm{MeV}$.
$\chi^2/\mathrm{DoF} = 0.9$ and CL = 33.2\%.
}
\label{fig:fig15} } }
\end{figure}

The logarithmic fit to the BW masses (MeV) of the five known $\Sigma 3/2^-$ excited states (blue circles) and four higher excited states projected (magenta diamonds) is shown in Fig.~\ref{fig:fig15}. $\Sigma3/2^-$ has a state reported in the PDG with a mass of $1675\pm 10~\mathrm{MeV}$, which does not fit Eq.~(\ref{eq:eq2}). We conclude that these masses are not well determined. In addition, two missed states (green triangles) are shown as calculated.
\subsection{Exotic Baryons}

\subsubsection{LHCb $P_{c\bar{c}}^+$ Excited-States}

A series of $P_{c\bar{c}}^+(?^?)$ exotic baryon resonances is recorded in the Particle Data Listings~\cite{ParticleDataGroup:2024cfk}. $P_{c\bar{c}}^+$: $I(J^P)S = 1/2(?^?)0$. These states were observed in the reaction $\Lambda_b^0\to P_{c\bar{c}}^+ K^- \to (J/\psi p) K^-$ by the LHCb Collaboration~\cite{LHCb:2019kea, LHCb:2015yax}. Since the logarithm equation fits the four data masses, their unknown $J^P$ values must be the same for the four data states.

The logarithmic fit to the BW masses (MeV) of the four known excited states $P_{c\bar{c}}^+(?^?)$ (blue circles) and the four highest excited states projected (magenta diamonds) is shown in
Fig.~15.  In addition, a missing state (green triangle) is shown as calculated.
\begin{figure}[htb!]
\centering
{
    \includegraphics[width=0.5\textwidth,keepaspectratio]{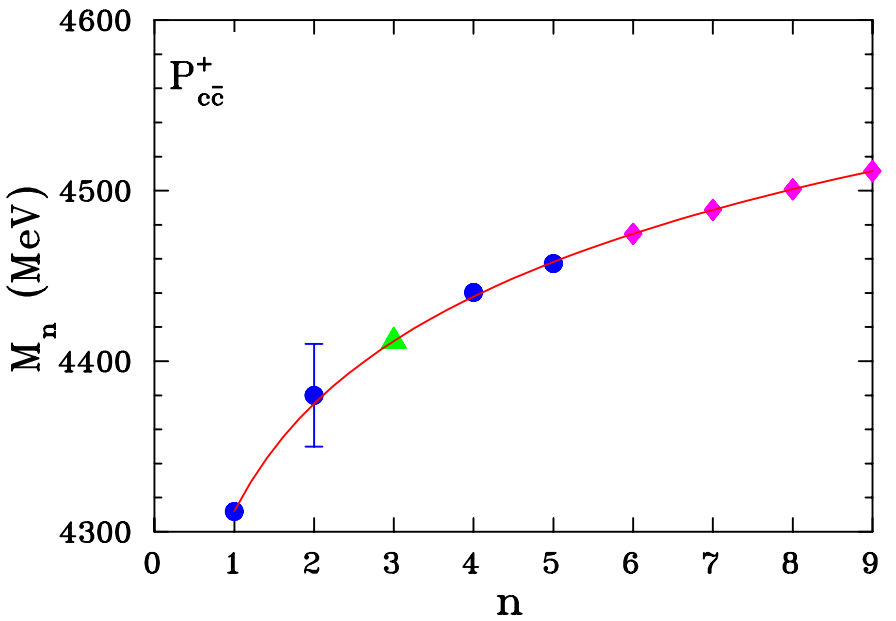} 
}

\centerline{\parbox{0.8\textwidth}{
\caption[] {\protect\small
Data for $P_{c\bar{c}}^+(?^?)$ (blue circles): $P_{c\bar{c}}(4312)^+$, $P_{c\bar{c}}(4380)^+$, $P_{c\bar{c}}(4440)^+$, and $P_{c\bar{c}}(4457)^+$~\cite{ParticleDataGroup:2024cfk}.
The green triangle is the calculated mass of the missing $P_{c\bar{c}}(4412)^+$ state with mass of $4411.8\pm 3.7~\mathrm{MeV}$. This state is weakly visitable but LHCb does not report it.
Predicted states (magenta diamonds): $P_{c\bar{c}}(4475)^+$, $P_{c\bar{c}}(4489)^+$, $P_{c\bar{c}}(4501)^+$, and $P_{c\bar{c}}(4511)^+$ with masses of $4474.7\pm 3.4~\mathrm{MeV}$, $4488.7\pm 3.3~\mathrm{MeV}$, $4500.8\pm 3.3~\mathrm{MeV}$, and $4511.5\pm 3.3~\mathrm{MeV}$, respectively.
The solid red curve presents the best-fit.
The fit parameters are $\alpha = 90.7\pm 2.9~\mathrm{MeV}$ and $\beta = 4312.2\pm 
3.8~\mathrm{MeV}$.
$\chi^2/\mathrm{DoF} = 0.2$ and CL = 82.5\%.}
}
\label{fig:fig17} } 
\end{figure}

\subsection{Cumulative Baryon Excited-States}

The cumulative fit curves of fifteen sets of baryons of equal quantities of mass are shown in Fig.~16.
\begin{figure}[htb!]
\centering
{
    \includegraphics[width=0.45\textwidth,keepaspectratio]{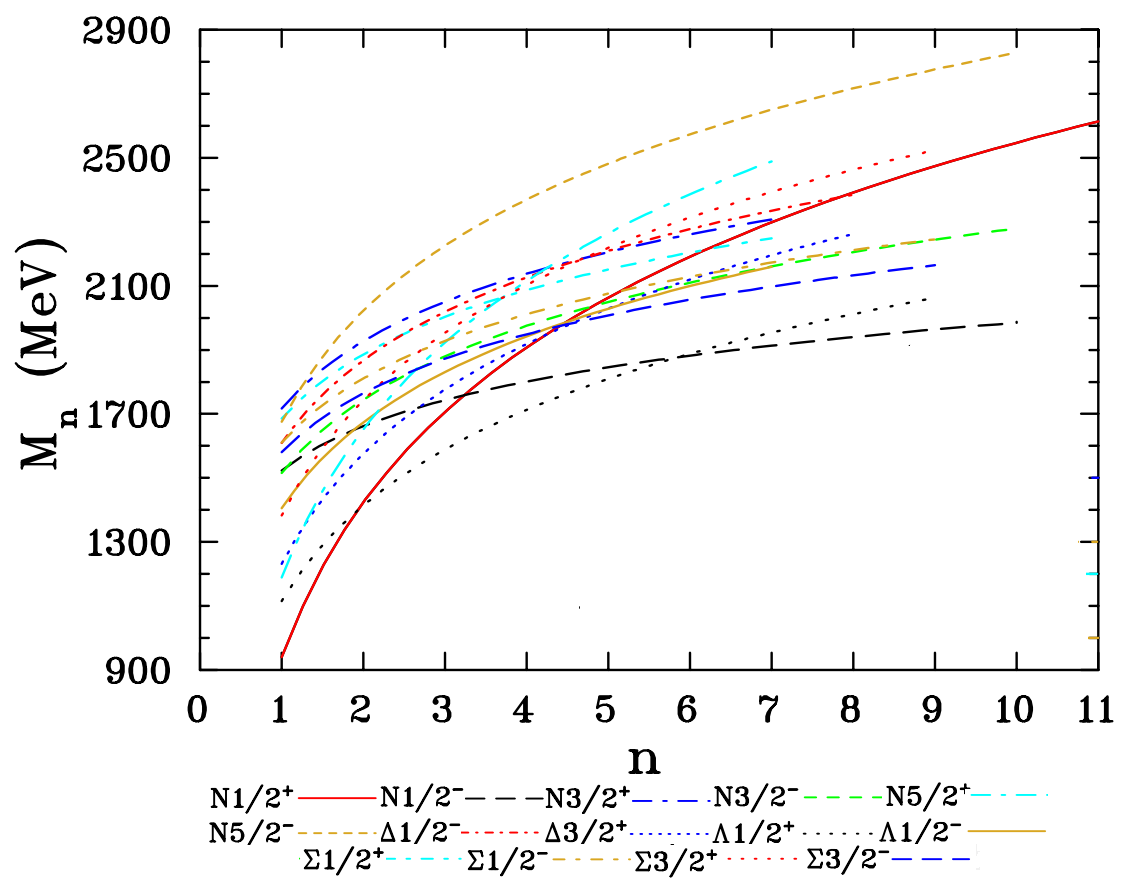}
    \includegraphics[width=0.47\textwidth,keepaspectratio]{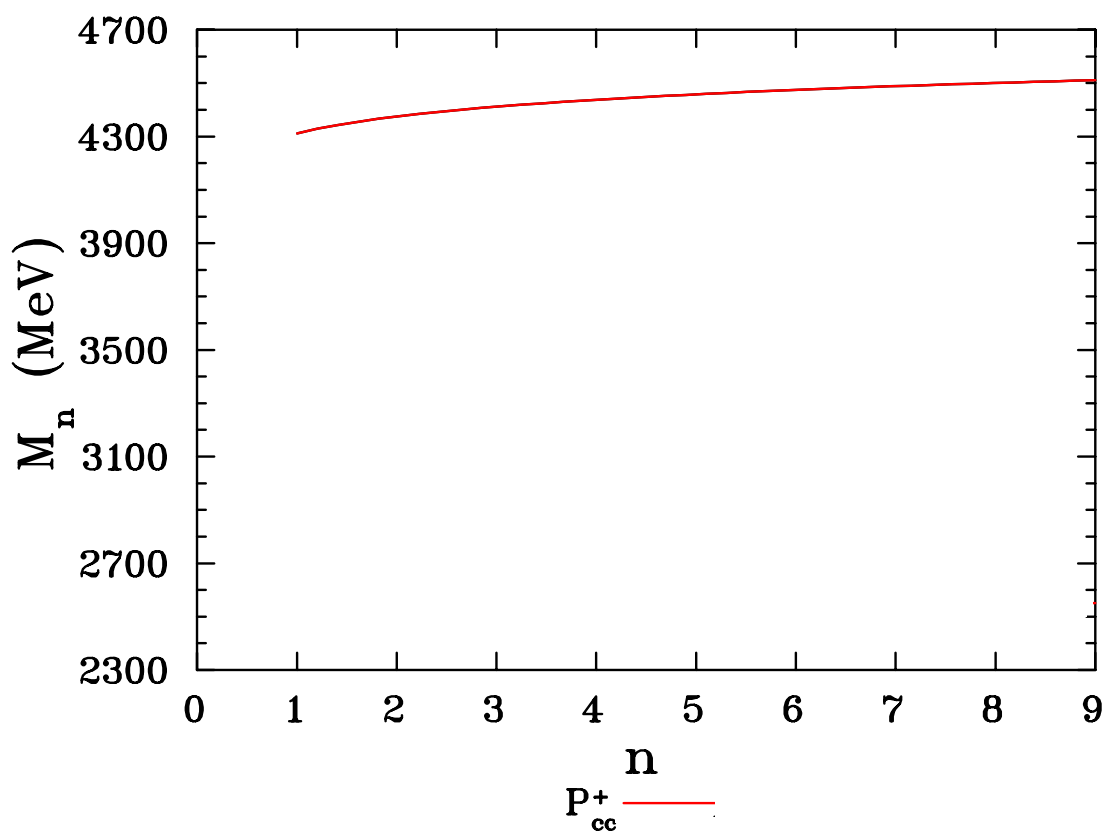}
}

\centerline{\parbox{0.8\textwidth}{
\caption[] {\protect\small
Cumulative fit curves for baryon excited states. 
\underline{Left}: $N1/2^+$, $N1/2^-$, $N3/2^+$, $N3/2^-$, $N5/2^+$, $N5/2^-$, $\Delta 1/2^-$, $\Delta 3/2^+$, $\Lambda 1/2^+$, $\Lambda 1/2^-$, 
$\Sigma 1/2^+$, $\Sigma 1/2^-$, $\Sigma 3/2^+$, and $\Sigma 3/2^-$.
\underline{Right}: $P_{c\bar{c}}^+$. 
Note that the ``logarithmic slope,'' $\alpha$ in Eq.~(\ref{eq:eq2}), usually decreases as the ground-state mass increases.
}}
\label{fig:bar} } 
\end{figure}

\subsection{PDG Star Rating}
The baryon sector of PDG~\cite{ParticleDataGroup:2024cfk} has a ``star'' rating system that clarifies the status of a particular resonance as
\begin{itemize}
\item $4\ast$  - Existence is certain.
\item $3\ast$  - Existence is very likely.
\item $2\ast$  - Evidence of existence is fair.
\item $1\ast$  - Evidence of existence is poor.   
\end{itemize}
We did not take this ``star'' rating into account. As an illustration, we fitted just $4\ast$ and $3\ast$ resonances to predict $2\ast$ and $1\ast$ cases to compare with PDG values.  Unfortunately, there are not many cases to do that, and we did as much as we could. So, here we report fits without a final low-star data point, instead predicting the mass of the low-star state:
\begin{itemize}
\item  $N1/2^+$ series has four $4\ast$ and one $3\ast$ low-laying states while a heavy state is $2\ast$. The previous fit (Subsec.~III.A.1) took into account $M = 2300\pm 65~\mathrm{MeV}$ with $\alpha = 698.2\pm 14.1~\mathrm{MeV}$ while the new fit gives $M = 2298\pm 60~\mathrm{MeV}$ with $\alpha = 667.3\pm 15.6~\mathrm{MeV}$.
\item  $N3/2^+$ series has two $4\ast$ low-laying states while a heavy state is $1\ast$. The previous fit (Subsec.~III.A.3) took into account $M = 2054\pm 27~\mathrm{MeV}$ with $\alpha = 304.4\pm 39.1~\mathrm{MeV}$ while the new fit gives $M = 2037\pm 48~\mathrm{MeV}$ with $\alpha = 288.5\pm 43.3~\mathrm{MeV}$.
\item  $N5/2^-$ series has one $4\ast$ and one $3\ast$ low-laying states while a heavy state is $2\ast$. The previous fit (Subsec.~III.A.5) took into account $M = 2570\pm 27~\mathrm{MeV}$ with $\alpha = 501.0\pm 15.6~\mathrm{MeV}$ while the new fit gives $M = 2573\pm 30~\mathrm{MeV}$ with $\alpha = 501.2\pm 15.0~\mathrm{MeV}$.
\item  $\Delta 1/2^-$ series has one $4\ast$ and one $3\ast$ low-laying states while a heavy state is $1\ast$. The previous fit (Subsec.~III.B.1) took into account $M = 2150\pm 100~\mathrm{MeV}$ with $\alpha = 373.0\pm 50.6~\mathrm{MeV}$ while the new fit gives $M = 2110\pm 80~\mathrm{MeV}$ with $\alpha = 360.7\pm 57.7~\mathrm{MeV}$
(Fig.~17).
\begin{figure}[htb!]
\centering
{
    \includegraphics[width=0.5\textwidth,keepaspectratio]{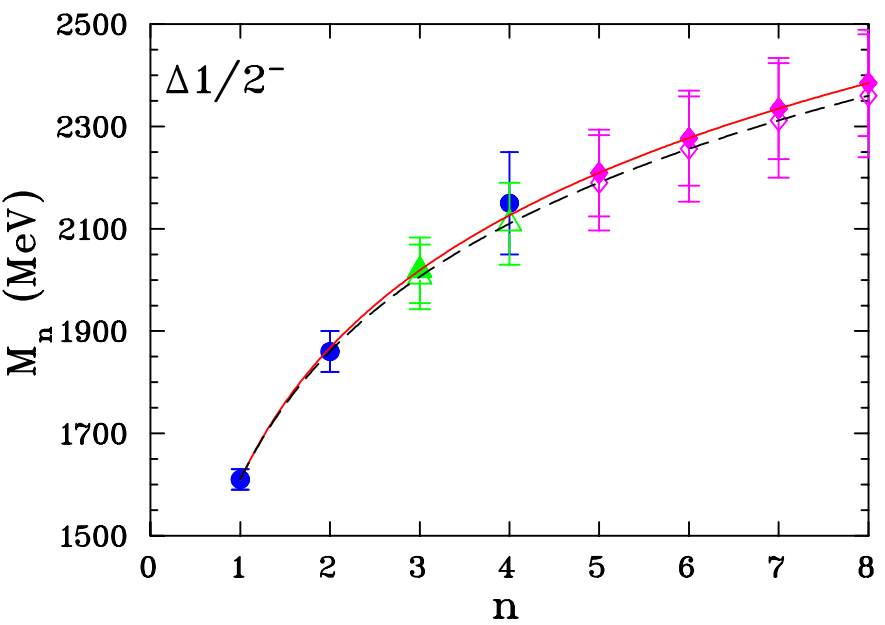} 
}

\centerline{\parbox{0.8\textwidth}{
\caption[] {\protect\small
Data for $\Delta 1/2^-$ (blue filled circles): $\Delta(1610)$ ($4\ast$), $\Delta(1860)$ ($3\ast$), and $\Delta(2150)$ ($1\ast$)~\cite{ParticleDataGroup:2024cfk}.
The green filled triangle is the calculated mass of the missing $\Delta(2016)$ state with mass  of $2019\pm 64~\mathrm{MeV}$.
Predicted states (magenta filled diamonds): $\Delta(2209)$, $\Delta(2277)$, $\Delta(2335)$, and $\Delta(2385)$.
Green open triangle is missed datum (star rating fit). 
Magenta open diamonds are predictions (star rating fit).
The solid red (dashed black) curve presents the best-fit (star rating fit).
For the star rating fit, we used just the two high-rate states and $\beta$ in Eq.~(\ref{eq:eq2}) was the ground state $\Delta(1610)$.}}
\label{fig:pdg} } 
\end{figure}
\item $\Lambda 1/2^+$ series has three $4\ast$ and one $3\ast$ states while one state is $1\ast$. The previous fit (Subsec.~III.C.1) took into account $M = 1713\pm 13~\mathrm{MeV}$ with $\alpha = 430.9\pm 8.5~\mathrm{MeV}$ while the new fit gives $M = 1714\pm 40~\mathrm{MeV}$ with $\alpha = 431.3\pm 20.5~\mathrm{MeV}$.
\item  $\Lambda 3/2^-$ series has two $4\ast$ states while one heavy state is $1\ast$. The previous fit (Subsec.~III.C.3) took into account $M = 2056\pm 22~\mathrm{MeV}$ with $\alpha = 250.4\pm 6.1~\mathrm{MeV}$ while the new fit gives $M = 2032\pm 15~\mathrm{MeV}$ with $\alpha = 246.7\pm 7.2~\mathrm{MeV}$.
\item  $\Sigma 1/2^+$ series has one $4\ast$ and one $3\ast$ states while one heavy state is $2\ast$. The previous fit (Subsec.~III.D.1) took into account $M = 1880\pm 60~\mathrm{MeV}$ with $\alpha = 668.0\pm 25.5~\mathrm{MeV}$ while the new fit gives $M = 1935\pm 32~\mathrm{MeV}$ with $\alpha = 679.0\pm 28.8~\mathrm{MeV}$.
\end{itemize}
This evaluation proves that our Universal Mass Equation (Eq.~(\ref{eq:eq2})) works well.

\section{Meson Excited-States}
\label{Sec:MES}

Sets of excited-meson states analyzed: Name($J^{PC}$)~[\# of Excited-States Data by PDG].\\
Mesons~(24) Data Sets): 
$\pi(0^{-+})$~[3], $\eta(0^{-+})$~[4], $\pi_2(2^{-+})$~[3], $\rho(1^{-~-})$~[4], $\rho_3(3^{-~-})$~[3], $f_0(0^{++})$~[9], $f_1(1^{++})$~[3], $f_2$(2$^{++}$)~[6], $\omega(1^{-~-})$~[4], $a_0(0^{++})$~[4], $h_1(1^{+-})$~[3],  $K(0^-)$~[3], $K_1(1^+)$~[3], $K_2(2^-)$~[3], $K^\ast_0(0^+)$~[3], $K^\ast(1^-)$~[3], $\phi(1^{-~-})$~[3], $\psi(1^{-~-})$~[6], $\chi_{c0}(0^{++})$~[4], $\chi_{c1}(1^{++})$~[4], $\Upsilon(1^{-~-})$~[7], $\chi_{b1}(1^{++})$~[3], $\chi_{b2}(2^{++})$~[3], and $T_{c\bar{c}1}(1^{+-})$~[3].

\subsection{Light Unflavored Mesons}

\subsubsection{$\pi(0^{-+})$ Excited-States}

A series of excited $\pi$-meson states is recorded in the Particle Data Listings~\cite{ParticleDataGroup:2024cfk}. $\pi$: $I^G(J^{PC}) = 
1^-(0^{-+})$ , $q\bar{q}$. 

\begin{figure}[htb!]
\centering
{
    \includegraphics[width=0.5\textwidth,keepaspectratio]{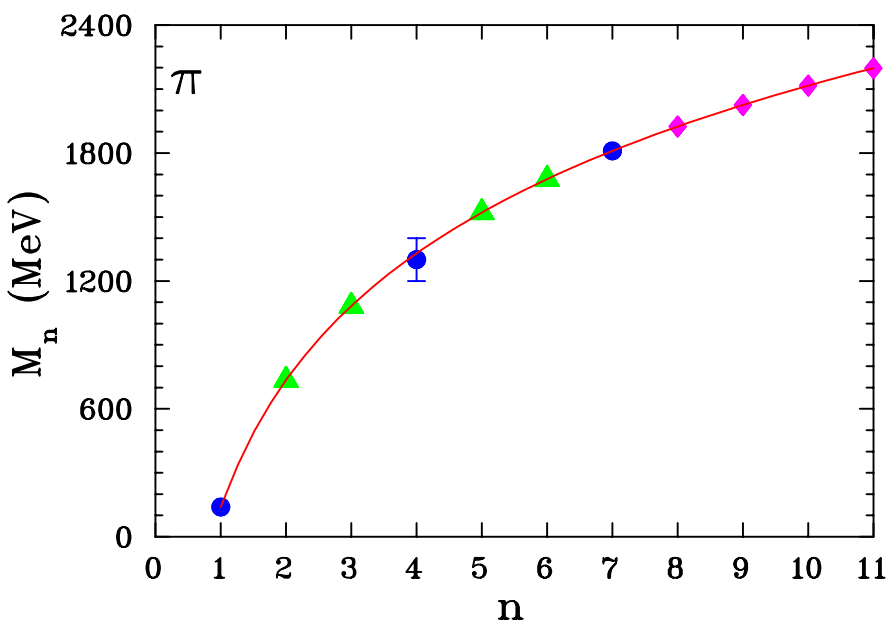} 
}

\centerline{\parbox{0.8\textwidth}{
\caption[] {\protect\small
Data for $\pi(0^{-+})$ (blue circles): $\pi(139)$, $\pi(1300)$, and $\pi(1810)$~\cite{ParticleDataGroup:2024cfk}.
The green triangles are the calculated masses of the missing $\pi(734)$, $\pi(1082)$, $\pi(1521)$, and $\pi(1677)$ states with masses of $734\pm 4~\mathrm{MeV}$, $1082\pm 6~\mathrm{MeV}$, $1521\pm 9~\mathrm{MeV}$, and $1677\pm 10~\mathrm{MeV}$, respectively.
Predicted states (magenta diamonds): $\pi(1924)$, $\pi(2025)$, $\pi(2116)$, and $\pi(2198)$ with masses of $1924\pm 11~\mathrm{MeV}$, $2025\pm 12~\mathrm{MeV}$, $2116\pm 12~\mathrm{MeV}$, and $2198\pm 13~\mathrm{MeV}$, respectively.
The solid red curve presents the best-fit.
The fit parameters are $\alpha = 858.3\pm 5.1~\mathrm{MeV}$ and $\beta = 139.6\pm 1.8\times 
10^{-4}~\mathrm{MeV}$.
$\chi^2/\mathrm{DoF} = 0.1$ and CL = 76.8\%.
}
\label{fig:fig19} } }
\end{figure}

The logarithmic fit to the BW masses (MeV) of the three known $\pi$-meson excited states (blue circles) and four projected higher excited states (magenta diamonds) is shown in Fig.~\ref{fig:fig19}.  In addition, four missing states (green triangles) are shown as calculated.

\subsubsection{$\eta(0^{-+})$ Excited-States}

A series of excited $\eta$-meson states is recorded in the Particle Data Listings~\cite{ParticleDataGroup:2024cfk}. $\eta$: $I^G(J^{PC}) = 
0^+(0^{-+})$ , $q\bar{q}$. 

The logarithmic fit to the BW masses (MeV) of the four known $\eta$-meson excited states (blue circles) and four projected higher excited states (magenta diamonds) is shown in Fig.~19.  $\eta$ has three states reported in the PDG with masses $1294\pm 4~\mathrm{MeV}$, $1408.7\pm 1.6~\mathrm{MeV}$, and $2221\pm 12~\mathrm{MeV}$, which do not fit Eq.~(\ref{eq:eq2}). We conclude that these masses are not well determined. 
PDG gives very small uncertainties for heavy-mass mesons $\eta(1476)$ and $\eta(1751)$ that do not allow 
Eq.~(\ref{eq:eq2}) to fit them well. We increased the uncertainties of the input data by a factor of 5 to present the results. 
In addition, four missing states (green triangles) are shown as calculated.
\begin{figure}[htb!]
\centering
{
    \includegraphics[width=0.5\textwidth,keepaspectratio]{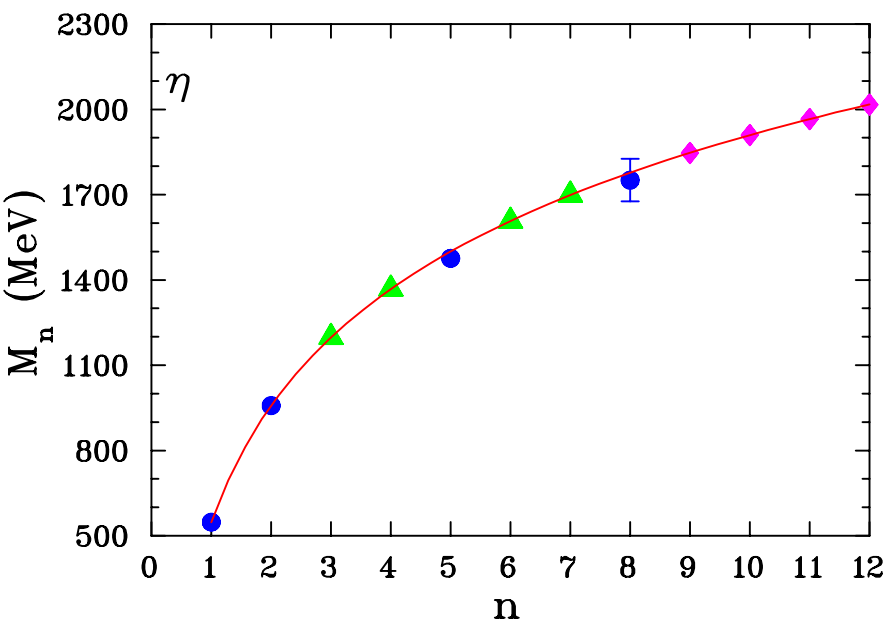} 
}

\centerline{\parbox{0.8\textwidth}{
\caption[] {\protect\small
Data for $\eta(0^{-+})$ (blue circles): $\eta(548)$, $\eta(958)$, $\eta(1476)$, and $\eta(1751)$~\cite{ParticleDataGroup:2024cfk}.
The green triangles are the calculated masses of the missing
$\eta(1198)$, $\eta(1368)$, $\eta(1607)$, and $\eta(1699)$ states with masses of $1198\pm 1~\mathrm{MeV}$, $1368\pm 1~\mathrm{MeV}$, $1607\pm 1~\mathrm{MeV}$, and $1699\pm 1~\mathrm{MeV}$, respectively.
Several states reported by PDG are out of the sequence $\eta(1294)$, $\eta(1409)$, and $\eta(2221)$.
Predicted states (magenta diamonds): $\eta(1847)$, $\eta(1910)$, $\eta(1966)$, and $\eta(2017)$ with masses of $1847\pm 1~\mathrm{MeV}$, $1910\pm 1~\mathrm{MeV}$, $1966\pm 1~\mathrm{MeV}$, and $2017\pm 1~\mathrm{MeV}$, respectively.
The solid red curve presents the best-fit.
The fit parameters are $\alpha = 591.39\pm 0.09~\mathrm{MeV}$ and 
$\beta = 547.86\pm 0.02~\mathrm{MeV}$.
$\chi^2/\mathrm{DoF} = 0.8$ and CL = 46.6\%.
}}
\label{fig:fig20} } 
\end{figure}

\subsubsection{$\pi_2(2^{-+})$ Excited-States}

A series of excited $\pi_2$-meson states is recorded in the Particle Data Listings~\cite{ParticleDataGroup:2024cfk}. $\pi_2$: $I^G(J^{PC}) = 1^-(2^{-+})$, $q\bar{q}$. 

\begin{figure}[htb!]
\centering
{
    \includegraphics[width=0.5\textwidth,keepaspectratio]{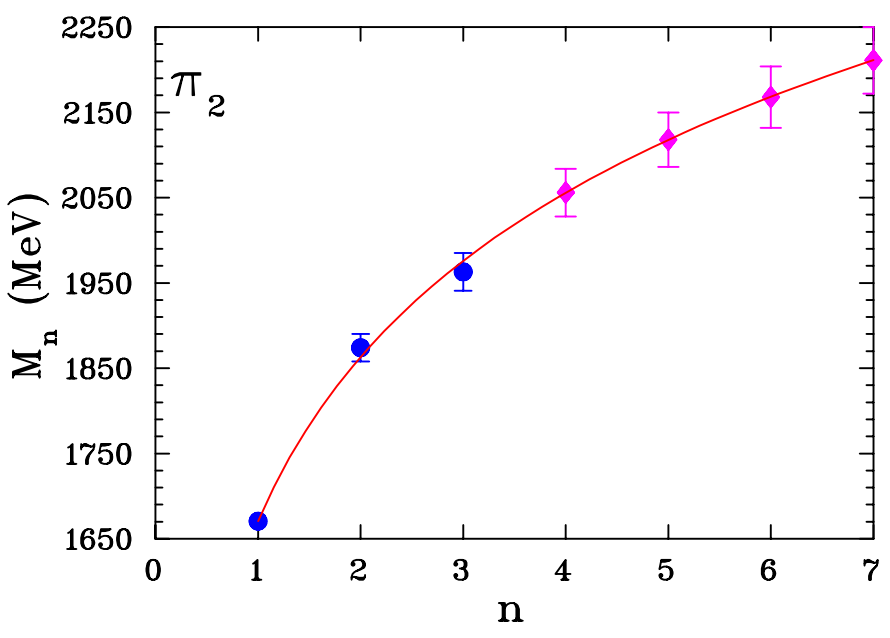} 
}

\centerline{\parbox{0.8\textwidth}{
\caption[] {\protect\small
Data for $\pi_2(2^{-+})$ (blue circles): $\pi_2(1671)$, $\pi_2(1874)$, and $\pi_2(1963)$~\cite{ParticleDataGroup:2024cfk}.
Predicted states (magenta diamonds): $\pi_2(2056)$, $\pi_2(2118)$, $\pi_2(2168)$, and $\pi_2(2211)$ with masses of $2056\pm 28~\mathrm{MeV}$, $2118\pm 32~\mathrm{MeV}$, $2168\pm 36~\mathrm{MeV}$, and $2211\pm 39~\mathrm{MeV}$, respectively.
The solid red curve presents the best-fit.
The fit parameters are $\alpha = 277.8\pm 15.3~\mathrm{MeV}$ and $\beta = 1670.7\pm 
2.1~\mathrm{MeV}$.
$\chi^2/\mathrm{DoF} = 0.8$ and CL = 37.2\%.
}
\label{fig:fig21} } }
\end{figure}

The logarithmic fit to the BW masses (MeV) of the three known $\pi_2$-meson excited states (blue circles) and four projected higher excited states (magenta diamonds) is shown in Fig.~\ref{fig:fig21}.

\subsubsection{$\rho(1^{-~-})$ Excited-States}

A series of excited $\rho$-meson states is recorded in the Particle Data Listings~\cite{ParticleDataGroup:2024cfk}. $\rho$: $I^G(J^{PC}) = 1^+(1^{-~-})$, $q\bar{q}$. 

\begin{figure}[htb!]
\centering
{
    \includegraphics[width=0.5\textwidth,keepaspectratio]{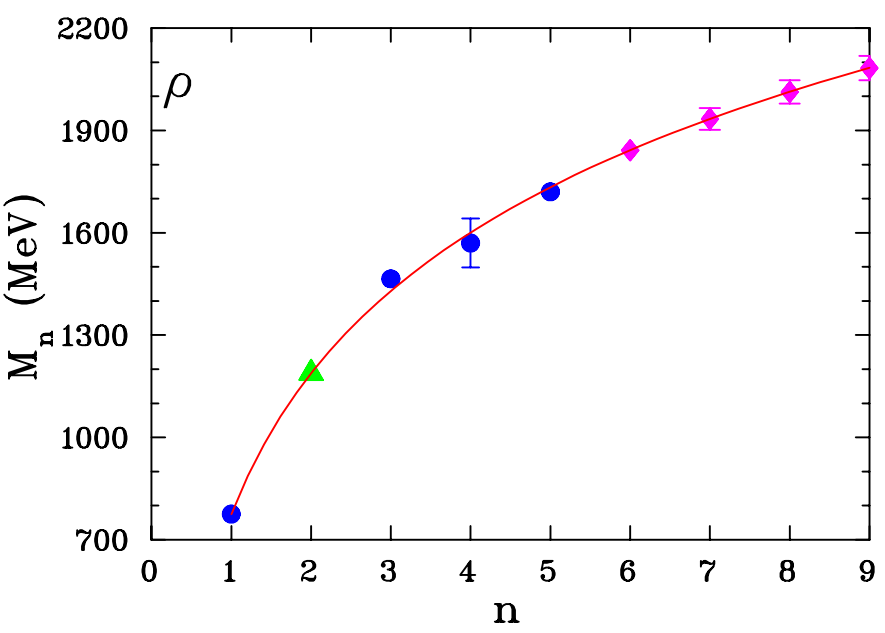} 
}

\centerline{\parbox{0.8\textwidth}{
\caption[] {\protect\small
Data for $\rho(1^{-~-})$ (blue circles): $\rho(775)$, $\rho(1465)$, $\rho(1570)$, and 
$\rho(1720)$~\cite{ParticleDataGroup:2024cfk}.
The green triangle is the calculated mass of the missing $\rho(1188)$ state with mass of $1188\pm 12~\mathrm{MeV}$.
Predicted states (magenta diamonds): $\rho(1842)$, $\rho(1934)$, $\rho(2013)$, and $\rho(2083)$ with masses of $1842\pm 29~\mathrm{MeV}$, $1934\pm 32~\mathrm{MeV}$, $2013\pm 34~\mathrm{MeV}$, and $2083\pm 36~\mathrm{MeV}$, respectively.
The solid red curve presents the best-fit.
The fit parameters are $\alpha = 595.4\pm 10.7~\mathrm{MeV}$ and $\beta = 775.3\pm 
0.2~\mathrm{MeV}$.
$\chi^2/\mathrm{DoF} = 1.3$ and CL = 26.3\%.
}
\label{fig:fig22} } }
\end{figure}

The logarithmic fit to the BW masses (MeV) of the four known $\rho$-meson excited states (blue circles) and one added missing state (green triangle) and four projected higher excited states (magenta diamonds) is shown in Fig.~\ref{fig:fig22}. In addition, a missing state (green triangle) is shown as calculated.

\subsubsection{$\rho_3(3^{-~-})$ Excited-States}

A series of excited $\rho_3$-meson states is recorded in the Particle Data Listings~\cite{ParticleDataGroup:2024cfk}. $\rho_3$: $I^G(J^{PC}) = 1^+(3^{-~-})$, $q\bar{q}$. 

\begin{figure}[htb!]
\centering
{
    \includegraphics[width=0.5\textwidth,keepaspectratio]{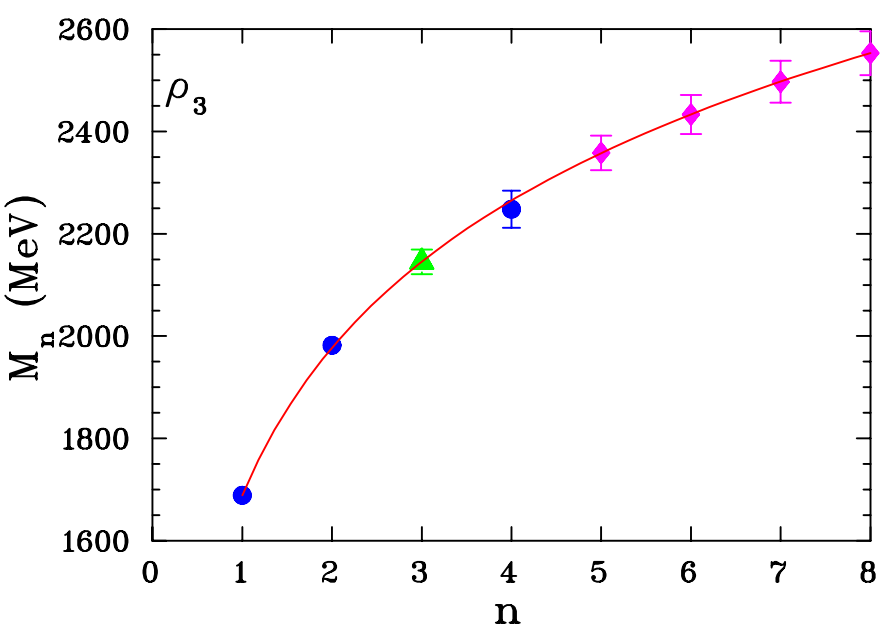} 
}

\centerline{\parbox{0.8\textwidth}{
\caption[] {\protect\small
Data for $\rho_3(3^{-~-})$ (blue circles): $\rho_3(1689)$, $\rho_3(1982)$, and 
$\rho_3(2248)$~\cite{ParticleDataGroup:2024cfk}.
The green triangle is the calculated mass of the missing $\rho_3(2145)$ state with mass of $2145\pm 24~\mathrm{MeV}$.
Predicted states (magenta diamonds): $\rho_3(2358)$, $\rho_3(2433)$, $\rho_3(2497)$, and $\rho_3(2553)$ with masses of $2358\pm 34~\mathrm{MeV}$, $2433\pm 38~\mathrm{MeV}$, $2497\pm 41~\mathrm{MeV}$, and $2553\pm 43~\mathrm{MeV}$, respectively.
The solid red curve presents the best-fit.
The fit parameters are $\alpha = 415.5\pm 16.1~\mathrm{MeV}$ and $\beta = 1688.9\pm 
2.1~\mathrm{MeV}$.
$\chi^2/\mathrm{DoF} = 0.4$ and CL = 55.1\%.
}
\label{fig:fig23} } }
\end{figure}

The logarithmic fit to the BW masses (MeV) of the four known $\rho_3$-meson excited states (blue circles) and four projected higher excited states (magenta diamonds) is shown in Fig.~\ref{fig:fig23}.  In addition, a missing state (green triangle) is shown as calculated.

\subsubsection{$f_0(0^{++})$ Excited-States}

A series of excited $f_0$-meson states is recorded in the Particle Data Listings at PDG~\cite{ParticleDataGroup:2024cfk}. $f_0$: $I^G(J^{PC}) = 0^+(0^{++})$, $q\bar{q}$. 

\begin{figure}[htb!]
\centering
{
    \includegraphics[width=0.5\textwidth,keepaspectratio]{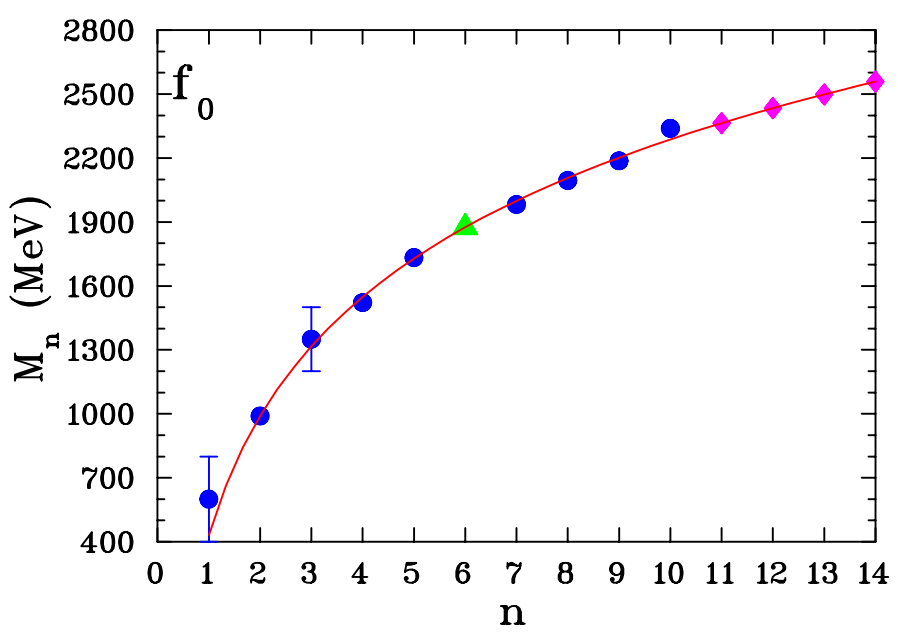} 
}

\centerline{\parbox{0.8\textwidth}{
\caption[] {\protect\small
Data for $f_0(0^{++})$ (blue circles): $f_0$(600), $f_0$(990), $f_0$(1350), $f_0$(1522), $f_0$(1733), $f_0$(1982), $f_0$(2095), $f_0$(2187), and $f_0$(2339)~\cite{ParticleDataGroup:2024cfk}.
The green triangle is the calculated mass of the missing $f_0$(1875) state with mass of $1875\pm 14~\mathrm{MeV}$.
Predicted states (magenta diamonds): $f_0$(2364), $f_0$(2434), $f_0$(2498), and $f_0$(2558) with masses of $2364\pm 13~\mathrm{MeV}$, $2434\pm 13~\mathrm{MeV}$, $2498\pm 13~\mathrm{MeV}$, and $2558\pm 13~\mathrm{MeV}$, respectively.
The solid red curve presents the best-fit. 
The fit parameters are $\alpha = 805.8\pm 13.6~\mathrm{MeV}$ and $\beta = 431.3\pm 
24.2~\mathrm{MeV}$.
$\chi^2/\mathrm{DoF} = 1.3$ and CL = 22.7\%.
}
\label{fig:fig24} } }
\end{figure}

The logarithmic fit to the BW masses (MeV) of the known $f_0$-meson excited states  (blue circles) and four projected higher excited states (magenta diamonds) is shown in Fig.~\ref{fig:fig24}. 
$f_0$ has two states reported in the PDG with masses $1784\pm 15~\mathrm{MeV}$ and $2470\pm 7~\mathrm{MeV}$, which do not fit Eq.~(\ref{eq:eq2}). We conclude that these masses are not well-determined.
In addition, two missing states (green triangles) are shown as calculated.

\subsubsection{$f_1(1^{++})$ Excited-States}

A series of excited $f_1$-meson states is recorded in the Particle Data Listings at PDG~\cite{ParticleDataGroup:2024cfk}. $f_1$: $I^G(J^{PC}) = 0^+(1^{++})$, $q\bar{q}$. 

\begin{figure}[htb!]
\centering
{
    \includegraphics[width=0.5\textwidth,keepaspectratio]{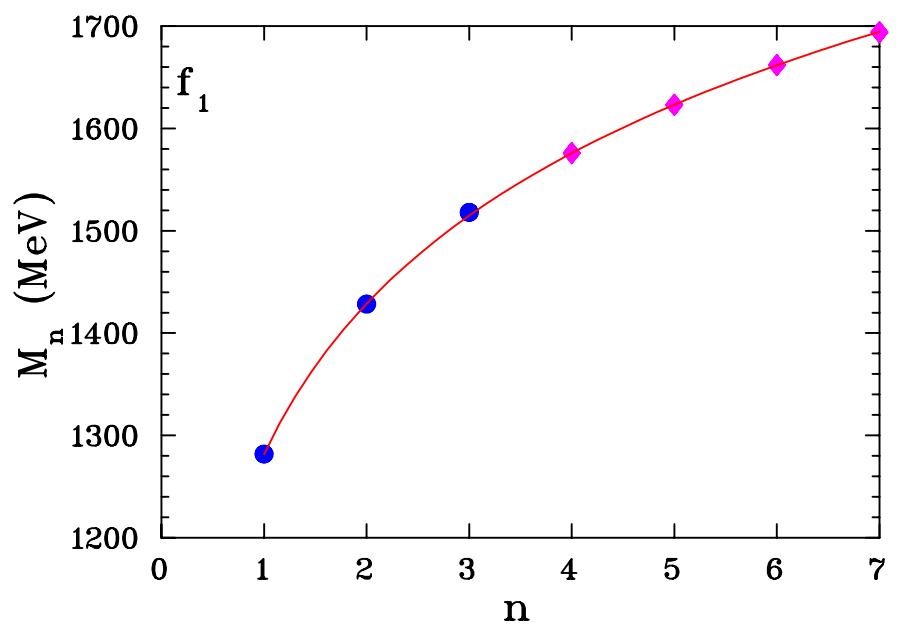} 
}

\centerline{\parbox{0.8\textwidth}{
\caption[] {\protect\small
Data for $f_1(1^{++})$ (blue circles): $f_1(1282)$, $f_1(1428)$, and $f_1(1518)$~\cite{ParticleDataGroup:2024cfk}.
Predicted states (magenta diamonds): $f_1(1576)$, $f_1(1623)$, $f_1(1662)$, and $f_1(1694)$ with masses of $1576\pm 3~\mathrm{MeV}$, $1623\pm 3~\mathrm{MeV}$, $1662\pm 4~\mathrm{MeV}$, and $1694\pm 4~\mathrm{MeV}$, respectively.
The solid red curve presents the best-fit.
The fit parameters are $\alpha = 210.3\pm 2.0~\mathrm{MeV}$ and $\beta = 1281.3\pm 0.5~\mathrm{MeV}$.
$\chi^2/\mathrm{DoF} = 0.2$ and CL = 88.5\%.
}
\label{fig:fig25} } }
\end{figure}

The logarithmic fit to the BW masses (MeV) of the known $f_1$-meson excited states  (blue circles) and four projected higher excited states (magenta diamonds) is shown in Fig.~\ref{fig:fig25}. 

\subsubsection{$f_2(2^{++})$ Excited-States}

A series of excited $f_2$-meson states is recorded in the Particle Data Listings~\cite{ParticleDataGroup:2024cfk}. $f_2$: $I^G(J^{PC}) = 
0^+(2^{++})$, $q\bar{q}$. 

\begin{figure}[htb!]
\vspace{-0.3cm}
\centering
{
    \includegraphics[width=0.5\textwidth,keepaspectratio]{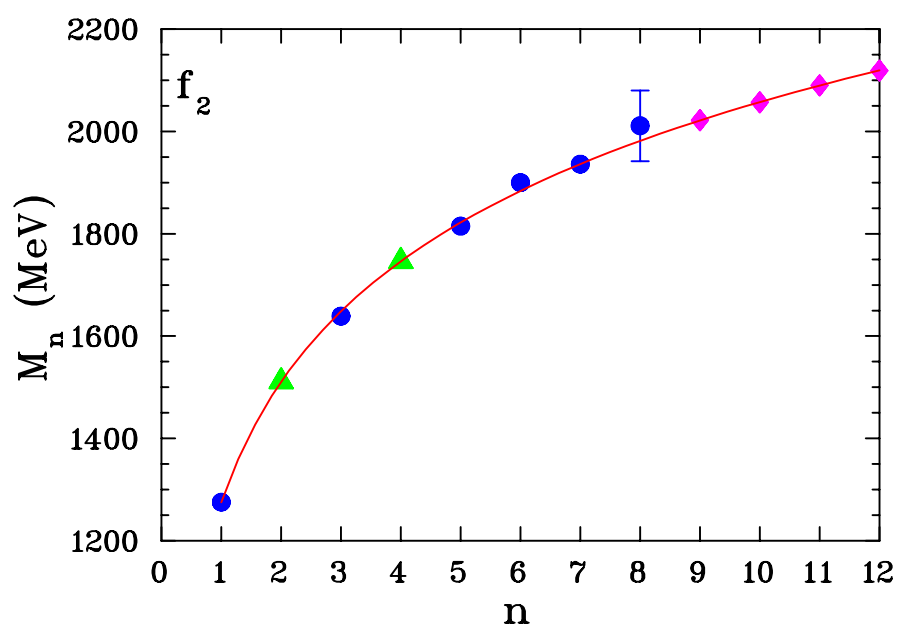} 
}

\centerline{\parbox{0.8\textwidth}{
\caption[] {\protect\small
Data for $f_2(2^{++})$ (blue circles): $f_2(1275)$, $f_2(1639)$, $f_2(1815)$, $f_2(1900)$, $f_2(1936)$, and 
$f_2$(2011)~\cite{ParticleDataGroup:2024cfk}.
The green triangles are the calculated masses of the missing $f_2(1511)$ and $f_2(1746)$ states with masses of $1511\pm 5~\mathrm{MeV}$ $1746\pm 4~\mathrm{MeV}$, respectively.
Several states reported by PDG are out of the sequence $f_2(1430)$, $f_2(1571)$, $f_2(2157)$, $f_2(2297)$, and $f_2(2346)$.
Predicted states (magenta diamonds): $f_2(2022)$, $f_2(2057)$, $f_2(2090)$, and $f_2(2119)$ 
The solid red curve presents the best-fit.
The fit parameters are $\alpha = 329.6\pm 2.9~\mathrm{MeV}$ and $\beta = 1275.3\pm 
0.8~\mathrm{MeV}$.
$\chi^2/\mathrm{DoF} = 1.5$ and CL = 18.3\%.
}
\label{fig:fig26} } }
\end{figure}

The logarithmic fit to the BW masses (MeV) of the known excited states $f_2$ (blue circles) and four projected higher excited states (magenta diamonds) is shown in Fig.~\ref{fig:fig26}. $f_2$ has five states reported in the PDG with masses $1430\pm 14~\mathrm{MeV}$, $1571\pm 13~\mathrm{MeV}$, $2157\pm 12~\mathrm{MeV}$, $2297\pm 28~\mathrm{MeV}$, and $2346\pm 16~\mathrm{MeV}$, which do not fit Eq.~(\ref{eq:eq2}). We conclude that these masses are not well determined. In addition, two missing states (green triangles) are shown as calculated.

\subsubsection{$\omega(1^{-~-})$ Excited-States}

A series of excited $\omega$-meson states is recorded in the Particle Data Listings~\cite{ParticleDataGroup:2024cfk}. $\omega$: 
$I^G(J^{PC}) = 0^-(1^{-~-})$, $q\bar{q}$. 

\begin{figure}[htb!]
\centering
{
    \includegraphics[width=0.5\textwidth,keepaspectratio]{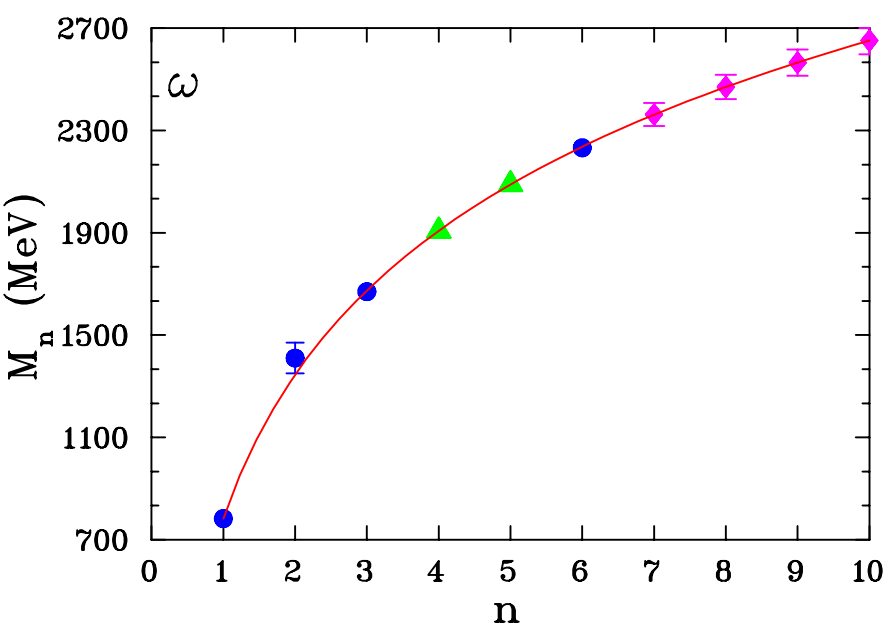} 
}

\centerline{\parbox{0.8\textwidth}{
\caption[] {\protect\small
Data for $\omega(1^{-~-})$ (blue circles): $\omega$(783), $\omega$(1410), $\omega$(1670), and 
$\omega$(2232)~\cite{ParticleDataGroup:2024cfk}.
The green triangles are the calculated masses of the missing $\omega(1907)$ and $\omega(2087)$ states with masses of $1907\pm 32~\mathrm{MeV}$ and $2089\pm 37~\mathrm{MeV}$, respectively.
Predicted states (magenta diamonds): $\omega$(2362), $\omega$(2470), $\omega$(2565), and $\omega$(2651) with masses of $2362\pm 45~\mathrm{MeV}$, $2470\pm 48~\mathrm{MeV}$, $2565\pm 51~\mathrm{MeV}$, and $2651\pm 53~\mathrm{MeV}$, respectively.
The solid red curve presents the best-fit.
The fit parameters are $\alpha = 811.4\pm 15.0~\mathrm{MeV}$ and $\beta = 782.7\pm 0.1~\mathrm{MeV}$.
$\chi^2/\mathrm{DoF} = 0.6$ and CL = 54.7\%.
}
\label{fig:fig27} } }
\end{figure}

The logarithmic fit to the BW masses (MeV) of the known $\omega$-meson excited states  (blue circles) and four projected higher excited states (magenta diamonds) is shown in Fig.~\ref{fig:fig27}. In addition, two missing states (green triangles) are shown as calculated.

\subsubsection{$a_0(0^{++})$ Excited-States}

A series of excited $a_0$-meson states is recorded in the Particle Data Listings~\cite{ParticleDataGroup:2024cfk}. $a_0$: $I^G~(J^{PC}) = 1^-(0^{++})$, $q\bar{q}$. 

\begin{figure}[htb!]
\centering
{
    \includegraphics[width=0.5\textwidth,keepaspectratio]{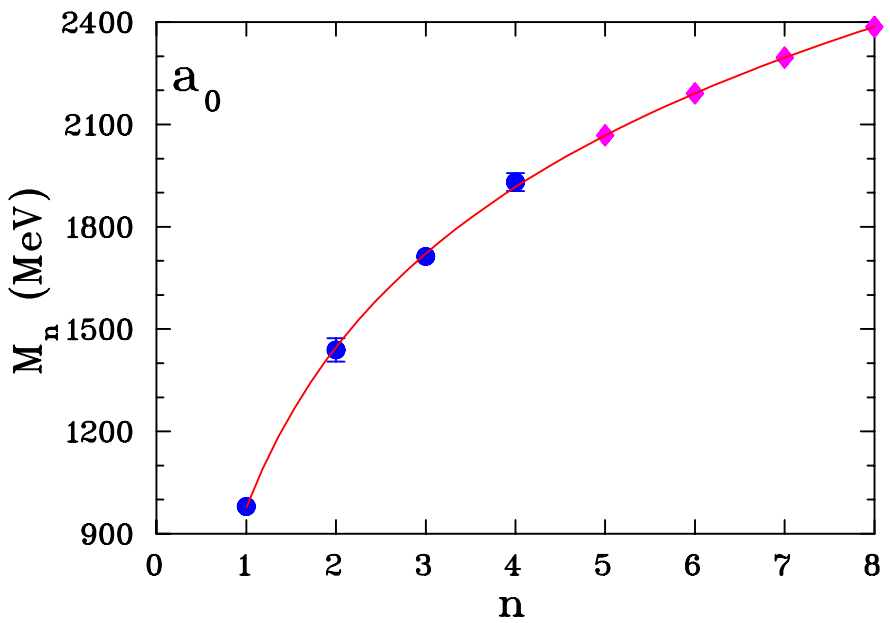} 
}

\centerline{\parbox{0.8\textwidth}{
\caption[] {\protect\small
Data for $a_0(0^{++})$ (blue circles): $a_0$(980), $a_0$(1439), $a_0$(1713), and $a_0$(1931)~\cite{ParticleDataGroup:2024cfk}.
Predicted states (magenta diamonds): $a_0$(2068), $a_0$(2191), $a_0$(2296), and $a_0$(2386) with masses of $2068\pm 24~\mathrm{MeV}$, $2191\pm 24~\mathrm{MeV}$, $2296\pm 24~\mathrm{MeV}$, and $2386\pm 25~\mathrm{MeV}$, respectively.
The solid red curve presents the best-fit.
The fit parameters are $\alpha = 677.8\pm 20.6~\mathrm{MeV}$ and $\beta = 976.8\pm 
19.2~\mathrm{MeV}$.
$\chi^2/\mathrm{DoF} = 0.3$ and CL = 74.6\%.
}
\label{fig:fig28} } }
\end{figure}

The logarithmic fit to the BW masses (MeV) of the four known $a_0$ -meson excited states (blue circles) and four projected higher excited states (magenta diamonds) is shown in Fig.~\ref{fig:fig28}.

\subsubsection{$h_1(1^{+-})$ Excited-States}

A series of excited $h_1$-meson states is recorded in the Particle Data Listings~\cite{ParticleDataGroup:2024cfk}. $h_1$: $I^G(J^{PC}) = 
0^-(1^{+~-})$, $q\bar{q}$. 

\begin{figure}[htb!]
\centering
{
    \includegraphics[width=0.5\textwidth,keepaspectratio]{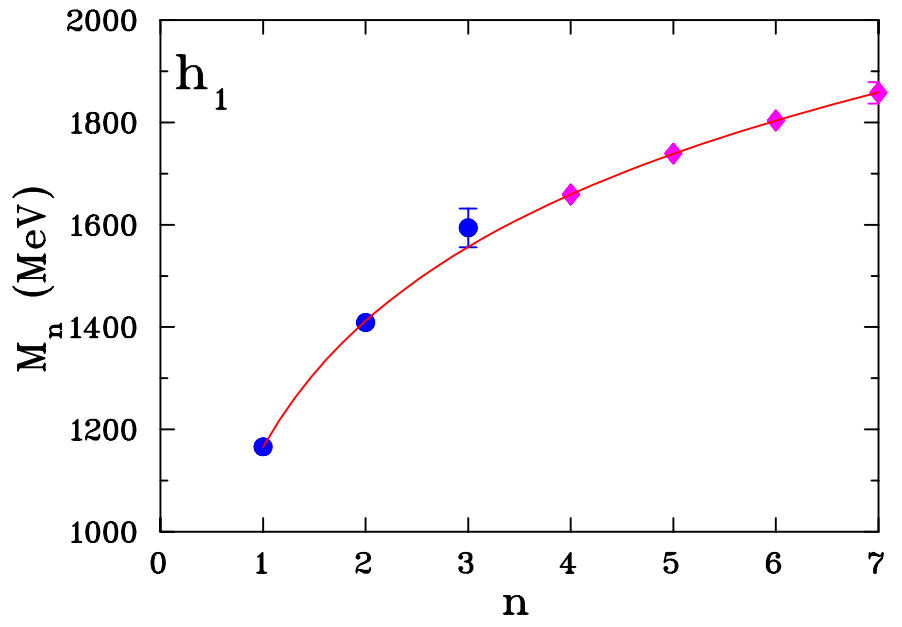} 
}

\centerline{\parbox{0.8\textwidth}{
\caption[] {\protect\small
Data for $h_1(1^{+-})$ (blue circles): $h_1$(1166), $h_1$(1409), and $h_1$(1594)~\cite{ParticleDataGroup:2024cfk}.
Predicted states (magenta diamonds): $h_1$(1659), $h_1$(1739), $h_1$(1804), and $h_1$(1858) with masses of $1659\pm 17~\mathrm{MeV}$, $1739\pm 18~\mathrm{MeV}$, $1804\pm 20~\mathrm{MeV}$, and $1858\pm 21~\mathrm{MeV}$, respectively.
The solid red curve presents the best-fit.
The fit parameters are $\alpha = 356.1\pm 14.7~\mathrm{MeV}$ and $\beta = 1165.5\pm 
6.0~\mathrm{MeV}$.
$\chi^2/\mathrm{DoF} = 1.1$ and CL = 29.3\%.
}
\label{fig:fig29} } }
\end{figure}

The logarithmic fit to the BW masses (MeV) of the three known $h_1$-meson excited states (blue circles) and four projected higher excited states (magenta diamonds) is shown in Fig.~\ref{fig:fig29}.

\subsection{Strange Mesons}

\subsubsection{$\phi(1^{-~-})$ Excited-States}

A series of excited $\phi$-meson states is recorded in the Particle Data Listings~\cite{ParticleDataGroup:2024cfk}. $\phi$: $I^G(J^{PC}) = 0^-(1^{-~-})$, $s\bar{s}$.  (Analog to $q\bar{q}\; \omega$ meson.)

\begin{figure}[htb!]
\centering
{
    \includegraphics[width=0.5\textwidth,keepaspectratio]{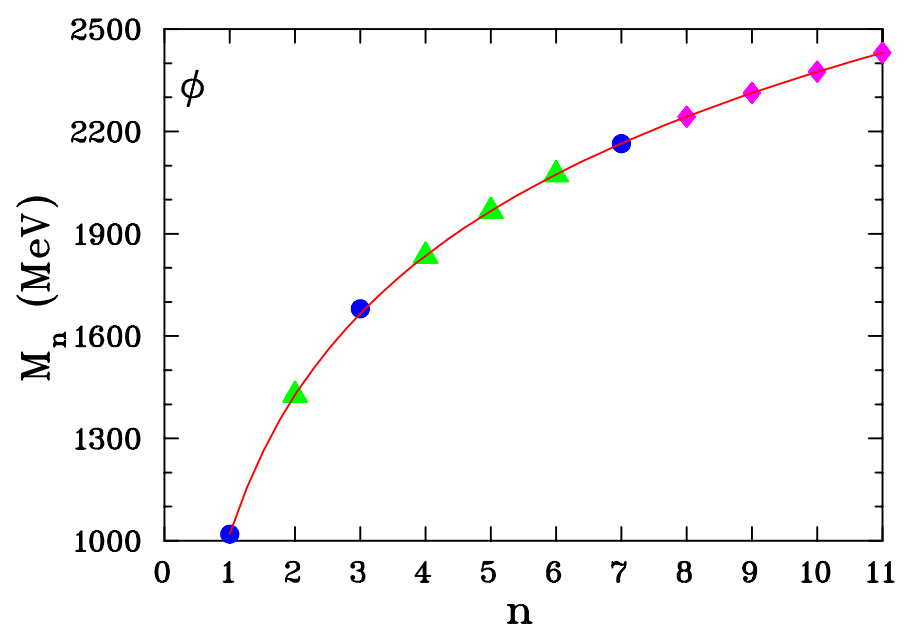} 
}

\centerline{\parbox{0.8\textwidth}{
\caption[] {\protect\small
Data for $\phi(1^{-~-})$ (blue circles): $\phi(1019)$, $\phi(1680)$, and $\phi(2164)$~\cite{ParticleDataGroup:2024cfk}.
The green triangles are the calculated masses of the missing $\phi(1427)$,  $\phi(1835)$, $\phi(1967)$, and $\phi(2074)$ states with masses of $1427\pm 3~\mathrm{MeV}$, $1835\pm 5~\mathrm{MeV}$, $1967\pm 5~\mathrm{MeV}$, and $2074\pm 6~\mathrm{MeV}$, respectively.
Predicted states (magenta diamonds): $\phi(2243)$, $\phi(2313)$, $\phi(2375)$, and $\phi(2431)$ with masses of $2243\pm 7~\mathrm{MeV}$, $2313\pm 8~\mathrm{MeV}$, $2375\pm 8~\mathrm{MeV}$, and $2431\pm 9~\mathrm{MeV}$, respectively.
The solid red curve presents the best-fit.
The fit parameters are $\alpha = 588.5\pm 3.0~\mathrm{MeV}$ and $\beta = 1019.50\pm 
0.02~\mathrm{MeV}$.
$\chi^2/\mathrm{DoF} = 0.5$ and CL = 47.9\%.
}
\label{fig:fig30} } }
\end{figure}

The logarithmic fit to the BW masses (MeV) of the three known $\phi$-meson excited states (blue circles) and four projected higher excited states (magenta diamonds) is shown in Fig.~\ref{fig:fig30}. In addition, four missing states (green triangles) are shown as calculated.

\subsubsection{$K(0^-)$ Excited-States}

A series of excited $K$-meson states is recorded in the Particle Data Listings~\cite{ParticleDataGroup:2024cfk}. $K$: $I(J^{PC}) = 
1/2(0^{-})$, $q\bar{s}$.

\begin{figure}[htb!]
\centering
{
    \includegraphics[width=0.5\textwidth,keepaspectratio]{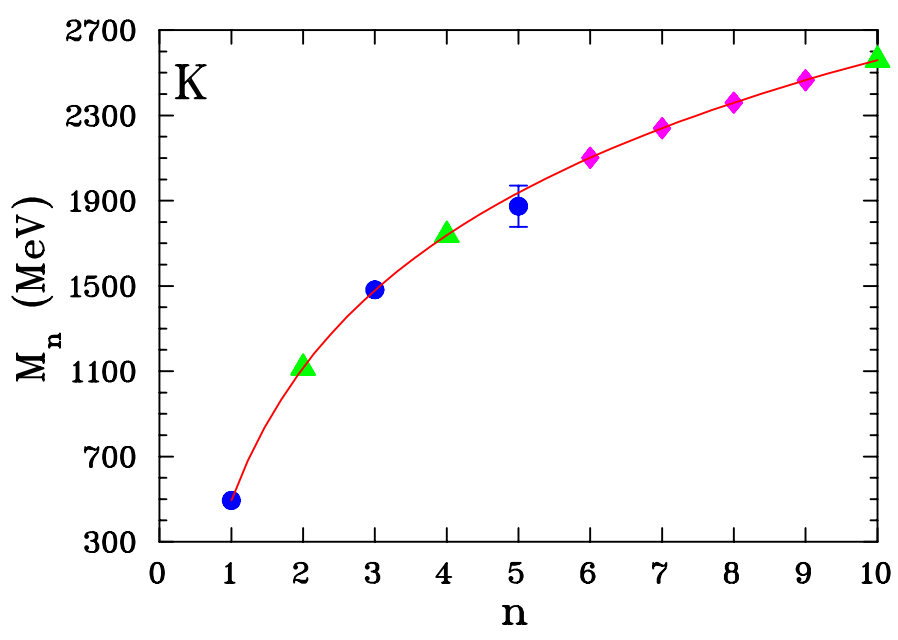} 
}

\centerline{\parbox{0.8\textwidth}{
\caption[] {\protect\small
Data for $K(0^-)$ (blue circles): $K(494)$, $K(1482)$, and 
$K(1874)$~\cite{ParticleDataGroup:2024cfk}.
The three green triangles are the calculated masses of the missing state $K(1116)$ and the two replacement states $K(1738)$ and $K(2560)$ with masses of $1116\pm 12~\mathrm{MeV}$, $1738\pm 24~\mathrm{MeV}$, and $2560\pm 39~\mathrm{MeV}$, respectively.
The predicted states (magenta diamonds) are $K(2101)$, $K(2240)$, $K(2360)$, and $K(2465)$.
The solid red curve presents the best-fit with masses of $2101\pm 31~\mathrm{MeV}$, $2240\pm 33~\mathrm{MeV}$, $2360\pm 35~\mathrm{MeV}$, and $2465\pm 37~\mathrm{MeV}$, respectively.
The fit parameters are $\alpha = 897.3\pm 14.2~\mathrm{MeV}$ and $\beta = 493.68\pm 0.02~\mathrm{MeV}$.
$\chi^2/\mathrm{DoF} = 0.5$ and CL = 49.8\%.
}
\label{fig:fig31} } }
\end{figure}

The logarithmic fit to the BW masses (MeV) of the three known $K$-meson excited states (blue circles) and three added missing states (green triangles) and four projected higher excited states (magenta diamonds) is shown in Fig.~\ref{fig:fig31}. $K(0^-)$ has two states reported in the PDG with masses $1629\pm 7~\mathrm{MeV}$ and $3054\pm 11~\mathrm{MeV}$ that do not fit Eq.~(\ref{eq:eq2}). We conclude that these masses are not well determined. Instead, we calculate their masses as $1734~\mathrm{MeV}$ and $2561~\mathrm{MeV}$, respectively. 

\subsubsection{$K_1(1^+)$ Excited-States}

A series of excited $K_1$-meson states is recorded in the Particle Data Listings~\cite{ParticleDataGroup:2024cfk}. $K_1$: $I(J^{PC}) = 
1/2(1^+)$, $q\bar{s}$.

\begin{figure}[htb!]
\centering
{
    \includegraphics[width=0.5\textwidth,keepaspectratio]{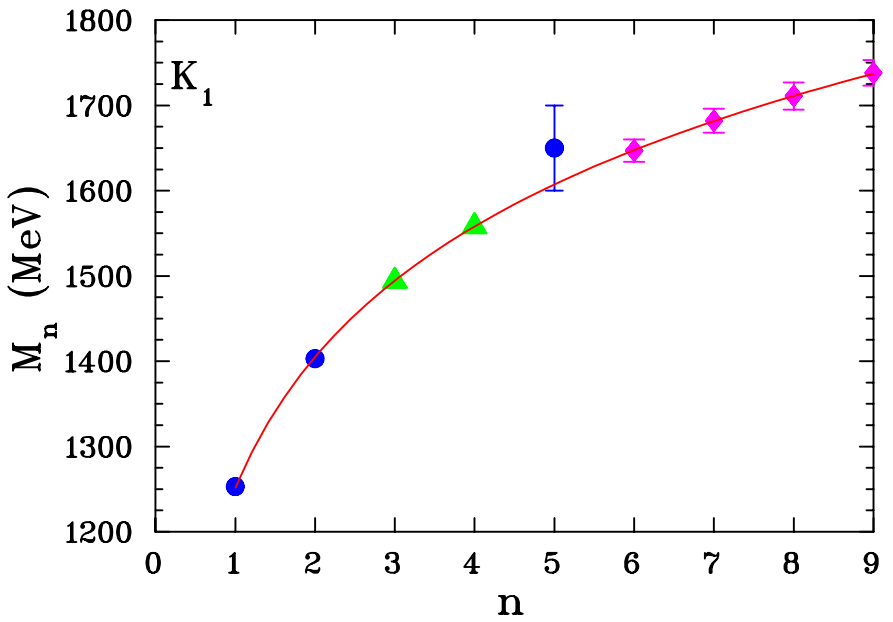} 
}

\centerline{\parbox{0.8\textwidth}{
\caption[] {\protect\small
Data for $K_1(1^+)$ (blue circles): $K_1(1253)$, $K_1(1403)$, and $K_1(1650)$~\cite{ParticleDataGroup:2024cfk}.
The two green triangles are the calculated masses of the missing states $K_1(1494)$ and $K_1(1558)$ with masses of $1494\pm 11~\mathrm{MeV}$ and $1558\pm 11~\mathrm{MeV}$, respectively.
The predicted states (magenta diamonds) are $K_1(1647)$, $K_1(1682)$, $K_1(1711)$, and $K_1(1738)$ with masses of $1647\pm 13~\mathrm{MeV}$, $1682\pm 14~\mathrm{MeV}$, $1711\pm 16~\mathrm{MeV}$, and $1737\pm 15~\mathrm{MeV}$, respectively.
The solid red curve presents the best-fit.
The fit parameters are $\alpha = 220.8\pm 13.4~\mathrm{MeV}$ and $\beta = 1251.9\pm 
6.9~\mathrm{MeV}$.
$\chi^2/\mathrm{DoF} = 0.8$ and CL = 36.1\%.
}
\label{fig:fig32} } }
\end{figure}

The logarithmic fit to the BW masses (MeV) of the three known $K_1$-meson excited states (blue circles) and two added missing states (green triangles) and four projected higher excited states (magenta diamonds) is shown in Fig.~\ref{fig:fig32}. In addition, two missing states (green triangles) are shown as calculated.

\subsubsection{$K_2(2^-)$ Excited-States}

A series of excited $K_2$-meson states is recorded in the Particle Data Listings~\cite{ParticleDataGroup:2024cfk}. $K_2$: $I(J^{PC}) = 
1/2(2^-)$, $q\bar{s}$.

\begin{figure}[htb!]
\centering
{
    \includegraphics[width=0.5\textwidth,keepaspectratio]{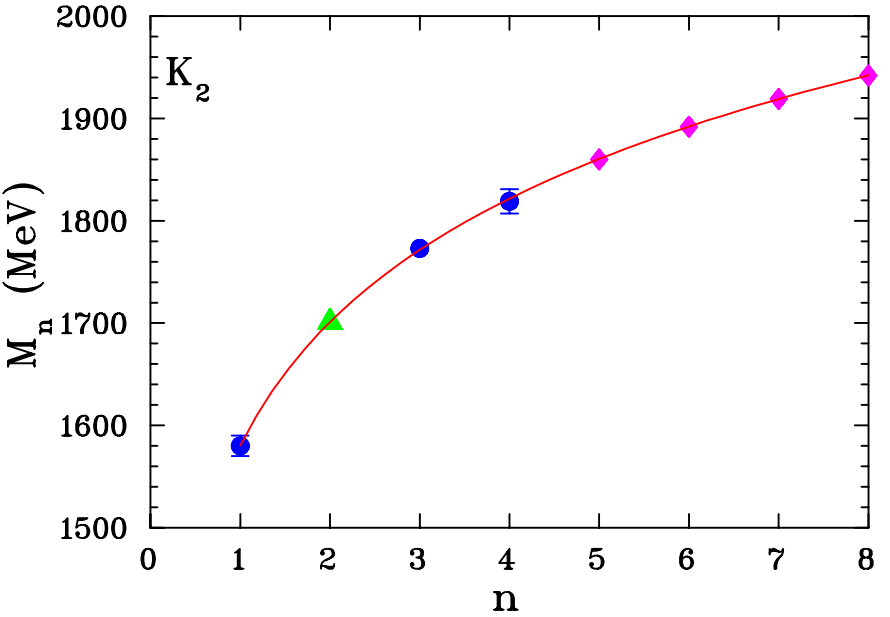} 
}

\centerline{\parbox{0.8\textwidth}{
\caption[] {\protect\small
Data for $K_2(2^-)$ (blue circles): $K_2(1580)$, $K_2(1773)$, and $K_2(1819)$~\cite{ParticleDataGroup:2024cfk}.
A state reported by PDG are out of the sequence $K_2(2247)$.
The green triangle is the calculated mass of the missing state $K_2(1701)$ with mass of $1701\pm 10~\mathrm{MeV}$.
The predicted states (magenta diamonds) are $K_2(1860)$, $K_2(1892)$, $K_2(1919)$, and $K_2(1942)$ with masses of $1860\pm 10~\mathrm{MeV}$, $1892\pm 9~\mathrm{MeV}$, $1919\pm 10~\mathrm{MeV}$, and $1942\pm 9~\mathrm{MeV}$, respectively.
The solid red curve presents the best-fit.
The fit parameters are $\alpha = 173.9\pm 9.9~\mathrm{MeV}$ and $\beta = 1580.5\pm 
9.9~\mathrm{MeV}$.
$\chi^2/\mathrm{DoF} = 0.1$ and CL = 77.5\%.
}
\label{fig:fig33} } }
\end{figure}

The logarithmic fit to the BW masses (MeV) of the three known $K_2$-meson excited states (blue circles) and four projected higher excited states (magenta diamonds) is shown in Fig.~\ref{fig:fig33}. $K_2(2^-)$ has a state reported in the PDG with mass $2247\pm 17~\mathrm{MeV}$ that does not fit Eq.~(\ref{eq:eq2}). We conclude that this mass is not well-determined. $K_2$ has a state reported in the PDG with a mass of $2247\pm 17~\mathrm{MeV}$, which does not fit Eq.~(\ref{eq:eq2}). We conclude that these masses are not well-determined.
In addition, a missing state (green triangle) is shown as calculated.

\subsubsection{$K^\ast_0(0^{+})$ Excited-States}

A series of excited $K^\ast_0$-meson states is recorded in the Particle Data Listings~\cite{ParticleDataGroup:2024cfk}. $K^\ast_0$: 
$I(J^{PC}) = 1/2(0^+)$, $q\bar{s}$. 

\begin{figure}[htb!]
\centering
{
    \includegraphics[width=0.5\textwidth,keepaspectratio]{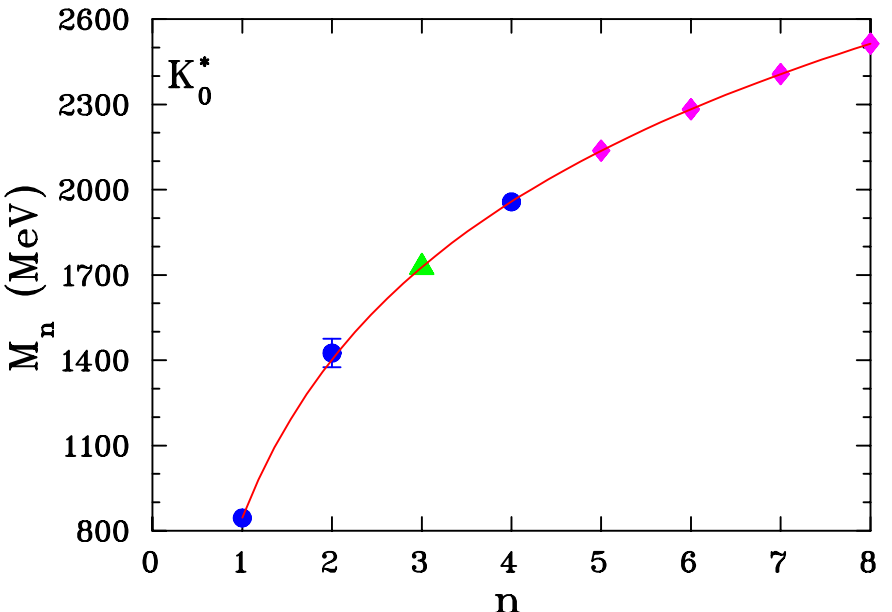} 
}

\centerline{\parbox{0.8\textwidth}{
\caption[] {\protect\small
Data for $K^\ast_0(0^+)$ (blue circles): $K^\ast_0(845)$, $K^\ast_0(1425)$, and $K^\ast_0(1957)$~\cite{ParticleDataGroup:2024cfk}.
The green triangle is the calculated mass of the missing state $K^\ast_0(1727)$ with mass of $1727\pm 16~\mathrm{MeV}$.
Predicted states (magenta diamonds): $K^\ast_0(2137)$, $K^\ast_0(2283)$, $K^\ast_0(2407)$, and $K^\ast_0(2514)$ with masses of $2137\pm 15~\mathrm{MeV}$, $2283\pm 14~\mathrm{MeV}$, $2407\pm 14~\mathrm{MeV}$, and $2514\pm 13~\mathrm{MeV}$, respectively.
The solid red curve presents the best-fit.
The fit parameters are $\alpha = 801.8\pm 15.9~\mathrm{MeV}$ and $\beta = 846.3\pm 
16.8~\mathrm{MeV}$.
$\chi^2/\mathrm{DoF} = 0.2$ and CL = 63.9\%.
}
\label{fig:fig34} } }
\end{figure}

The logarithmic fit to the BW masses (MeV) of the three known $K^\ast_0$-meson excited states (blue circles) and four projected higher excited states (magenta diamonds) is shown in Fig.~\ref{fig:fig34}. In addition, a missing state (green triangle) is shown as calculated.

\subsubsection{$K^\ast(1^{-})$ Excited-States}

A series of excited $K^\ast$-meson states is recorded in the Particle Data Listings~\cite{ParticleDataGroup:2024cfk}. $K^\ast$: 
$I(J^{PC}) = 1/2(1^-)$, $q\bar{s}$. 

\begin{figure}[htb!]
\centering
{
    \includegraphics[width=0.5\textwidth,keepaspectratio]{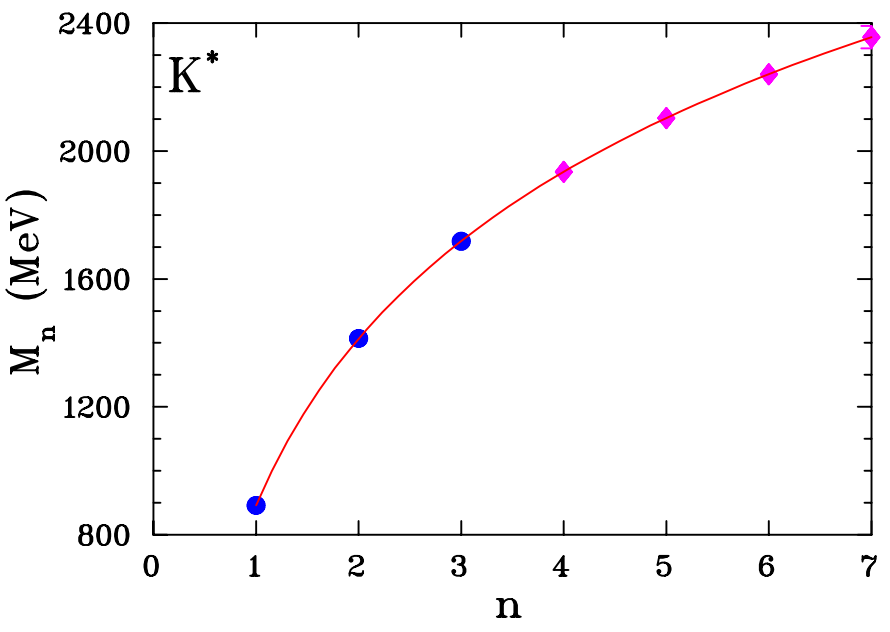} 
}

\centerline{\parbox{0.8\textwidth}{
\caption[] {\protect\small
Data for $K^\ast(1^-)$ (blue circles): $K^\ast$(892), $K^\ast$(1414), and $K^\ast$(1718)~\cite{ParticleDataGroup:2024cfk}.
Predicted states (magenta diamonds): $K^\ast$(1935), $K^\ast$(2103), $K^\ast$(2240), and $K^\ast$(2356) with masses of $1935\pm 25~\mathrm{MeV}$, $2103\pm 29~\mathrm{MeV}$, $2240\pm 32~\mathrm{MeV}$, and $2356\pm 35~\mathrm{MeV}$, respectively.
The solid red curve presents the best-fit.
The fit parameters are $\alpha = 752.7\pm 13.1~\mathrm{MeV}$ and $\beta = 891.7\pm 
0.3~\mathrm{MeV}$.
$\chi^2/\mathrm{DoF} = 0.003$ and CL = 95.9\%.
}
\label{fig:fig35} } }
\end{figure}

The logarithmic fit to the BW masses (MeV) of the three known $K^\ast$-meson excited states (blue circles) and four projected higher excited states (magenta diamonds) is shown in Fig.~\ref{fig:fig35}.

\subsection{$c\bar{c}$ Mesons}

\subsubsection{$\psi(1^{-~-})$ Excited-States}

A series of excited $\psi$-meson states is recorded in the Particle Data Listings~\cite{ParticleDataGroup:2024cfk}. $\psi$: 
$I^G(J^{PC}) = 0^-(1^{-~-})$, $c\bar{c}$.  (Analog to $q\bar{q}\;\omega$ meson.)

\begin{figure}[htb!]
\centering
{
    \includegraphics[width=0.5\textwidth,keepaspectratio]{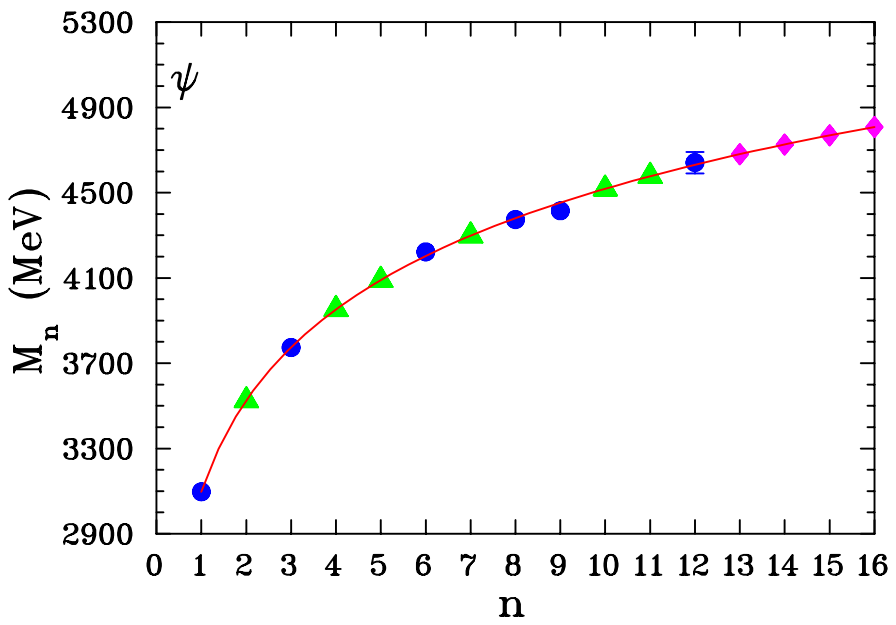} 
}

\centerline{\parbox{0.8\textwidth}{
\caption[] {\protect\small
Data for $\psi(1^{-~-})$ (blue circles): $J/\psi(3097)$, $\psi(3774)$, $\psi(4222)$, $\psi(4374)$, $\psi(4415)$, and  $\psi(4641)$~\cite{ParticleDataGroup:2024cfk}. The green triangles are the calculated masses of the missing states $\psi(3525)$, $\psi(3953)$, $\psi(4090)$, $\psi(4298)$, $\psi(4518)$, and $\psi(4577)$ with masses of $3525\pm 3~\mathrm{MeV}$, $3953\pm 7~\mathrm{MeV}$, $4090\pm 8~\mathrm{MeV}$, $4298\pm 9~\mathrm{MeV}$, $4518\pm 11~\mathrm{MeV}$, and $4577\pm 12~\mathrm{MeV}$, respectively.
Predicted states (magenta diamonds): $\psi(4680)$, $\psi(4726)$, $\psi(4769)$, and $\psi(4808)$ with masses of $4680\pm 12~\mathrm{MeV}$, $4726\pm 13~\mathrm{MeV}$, $4769\pm 13~\mathrm{MeV}$, and $4808\pm 14~\mathrm{MeV}$, respectively.
The solid red curve presents the best-fit.
The fit parameters are $\alpha = 617.3\pm 2.7~\mathrm{MeV}$ and $\beta = 3096.90\pm 0.03~\mathrm{MeV}$.
$\chi^2/\mathrm{DoF} = 1.3$ and CL = 25.4\%, taking into account the fact that we increased uncertainties of input data by a factor of 5.
}
\label{fig:fig36} } }
\end{figure}

The logarithmic fit to the BW masses (MeV) of the three known $\psi$-meson excited states (blue circles) and four projected higher excited states (magenta diamonds) is shown in Fig.~\ref{fig:fig36}. $\psi$ has three states reported in the PDG with masses $\psi(3686)$, $\psi(4040)$, and $\psi(4191)$ that do not fit Eq.~(\ref{eq:eq2}). We conclude that these masses are not well-determined.
PDG gives very small uncertainties for heavy-mass mesons of approximately 0.2\% that do not allow Eq.~(\ref{eq:eq2}) to 
fit them well. We increased the uncertainties of the input data by a factor of 5 to present the results.
In addition, the six missing states (green triangles) are shown as calculated.

\subsubsection{$\chi_c(0) (0^{++})$ Excited-States}

A series of excited $\chi_c$-meson states is recorded in the Particle Data Listings~\cite{ParticleDataGroup:2024cfk}. $\chi_c(0)$: $I^G(J^{PC}) = 0^+(0^{++})$, $c\bar{c}$.  (Analog to $q\bar{q}\;f_0$ meson.)

The logarithmic fit to the BW masses (MeV) of the four known $\chi_c(0)$-meson excited states (blue circles) and four projected higher excited states (magenta diamonds) is shown in Fig.~36. $\chi_c(0)$ has a state reported in the PDG with mass $\chi_c(0)(3922)$ that does not fit Eq.~(\ref{eq:eq2}). We conclude that these masses are not well determined.
PDG gives very small uncertainties for heavy mass mesons of approximately 0.05\% that do not allow Eq.~(\ref{eq:eq2}) to 
fit them well. We increased the uncertainties of the input data by a factor of 5 to present the results.
In addition, the missing state (green triangle) is shown as calculated.
\begin{figure}[htb!]
\centering
{
    \includegraphics[width=0.5\textwidth,keepaspectratio]{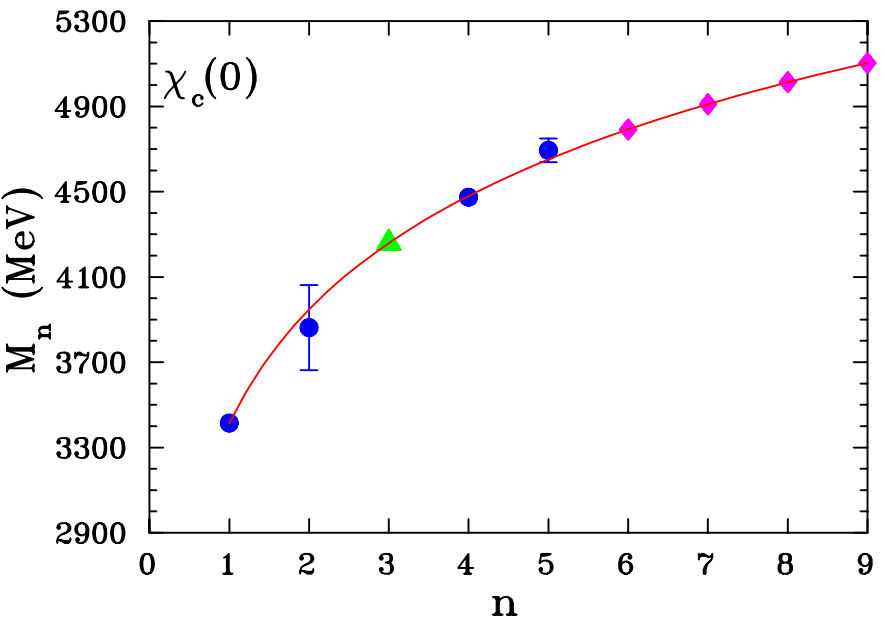} 
}

\centerline{\parbox{0.8\textwidth}{
\caption[] {\protect\small
Data for $\chi_c(0)(0^{++})$ (blue circles): $\chi_c(0)(3415)$, $\chi_c(0)(3862)$, $\chi_c(0)(4474)$, and  $\chi_c(0)(4694)$~\cite{ParticleDataGroup:2024cfk}. The green triangle is the calculated mass of the missing state $\chi_c(0)(4259)$ with mass of $4259\pm 20~\mathrm{MeV}$.
Predicted states (magenta diamonds): $\chi_c(0)(4792)$, $\chi_c(0)(4910)$, $\chi_c(0)(5013)$, and $\chi_c(0)(5103)$ with masses of $4792\pm 32~\mathrm{MeV}$, $4910\pm 34~\mathrm{MeV}$, $5013\pm 37~\mathrm{MeV}$, and $5103\pm 39~\mathrm{MeV}$, respectively.
The solid red curve presents the best-fit.
The fit parameters are $\alpha = 768.5\pm 13.3~\mathrm{MeV}$ and $\beta = 3414.7\pm 
1.5~\mathrm{MeV}$.
$\chi^2/\mathrm{DoF} = 0.4$ and CL = 64.7\%, taking into account the fact that we increased uncertainties of input data by a factor of 5.
}}
\label{fig:fig37} } 
\end{figure}

\subsubsection{$\chi_c(1)(1^{++})$ Excited-States}

A series of excited $\chi_c(1)$-meson states is recorded in the Particle Data Listings~\cite{ParticleDataGroup:2024cfk}. $\chi_c(1)$: $I^G(J^{PC}) = 0^+(1^{++})$, $c\bar{c}$.  (Analog to the $q \bar{q}\; f_1$ meson.)

The logarithmic fit to the BW masses (MeV) of the four known $\chi_c(1)$-meson excited states (blue circles) and four projected higher excited states (magenta diamonds) is shown in Fig.~37. $\chi_c(1)$ has a state reported in the PDG with mass $\chi_c(1)(3872)$ that does not fit Eq.~(\ref{eq:eq2}). We conclude that these masses are not well-determined.
PDG gives very small uncertainties for heavy-mass mesons of approximately 0.002\% that do not allow Eq.~(\ref{eq:eq2}) to 
fit them well. We increased the uncertainties of the input data by a factor of 2 to present the results.
In addition, four missing states (green triangles) are shown as calculated.
\begin{figure}[htb!]
\centering
{
    \includegraphics[width=0.5\textwidth,keepaspectratio]{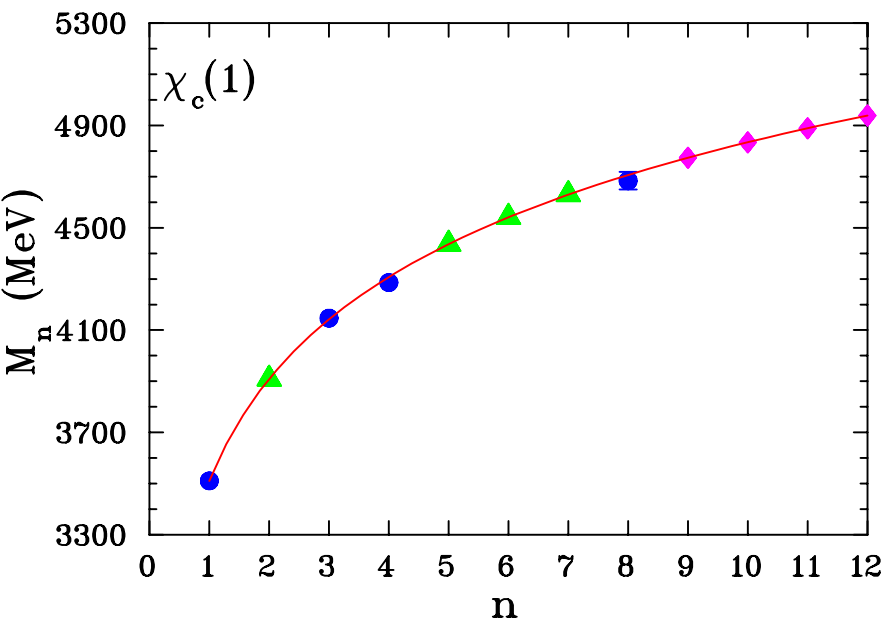} 
}

\centerline{\parbox{0.8\textwidth}{
\caption[] {\protect\small
Data for $\chi_c(1)(1^{++})$ (blue circles): $\chi_c(1)(3511)$, $\chi_c(1)(4146)$, $\chi_c(1)(4286)$, and  
$\chi_c(1)(4684)$~\cite{ParticleDataGroup:2024cfk}. The green triangles are the calculated mass of the missing states $\chi_c(1)(3909)$, $\chi_c(1)(4436)$, $\chi_c(1)(4541)$, and $\chi_c(1)(4630)$ with masses of $3909\pm 5~\mathrm{MeV}$, $4436\pm 12~\mathrm{MeV}$, $4541\pm 13~\mathrm{MeV}$, and $4629\pm 14~\mathrm{MeV}$, respectively.
Predicted states (magenta diamonds): $\chi_c(1)(4774)$, $\chi_c(1)(4834)$, $\chi_c(1)(4889)$, and $\chi_c(1)(4939)$ with masses of $4774\pm 16~\mathrm{MeV}$, $4834\pm 17~\mathrm{MeV}$, $4889\pm 18~\mathrm{MeV}$, and $4939\pm 18~\mathrm{MeV}$, respectively.
The solid red curve presents the best-fit.
The fit parameters are $\alpha = 574.8\pm 4.8~\mathrm{MeV}$ and $\beta = 3510.7\pm 0.1~\mathrm{MeV}$.
$\chi^2/\mathrm{DoF} = 1.2$ and CL = 30.6\%, taking into account the fact that we increased uncertainties of input data by a factor of 2.
}}
\label{fig:fig38} } 
\end{figure}

\subsubsection{$T_{c\bar{c}1}(1^{+-})$ Excited-States}

A series of excited $T_{c\bar{c}1}(1^{+-})$-meson states is recorded in the Particle Data Listings~\cite{ParticleDataGroup:2024cfk}. $T_{c\bar{c}1}(1^{+-})$: 
$I^G(J^{PC}) = 1^+(1^{+-})$, $c\bar{c}$.  (Analog to the $q\bar{q}\;b_1$ meson.)

The logarithmic fit to the BW masses (MeV) of the four known $T_{c\bar{c}1}$-meson excited states (blue circles) and four projected higher excited states (magenta diamonds) is shown in Fig.~38. 
\begin{figure}[htb!]
\centering
{
    \includegraphics[width=0.5\textwidth,keepaspectratio]{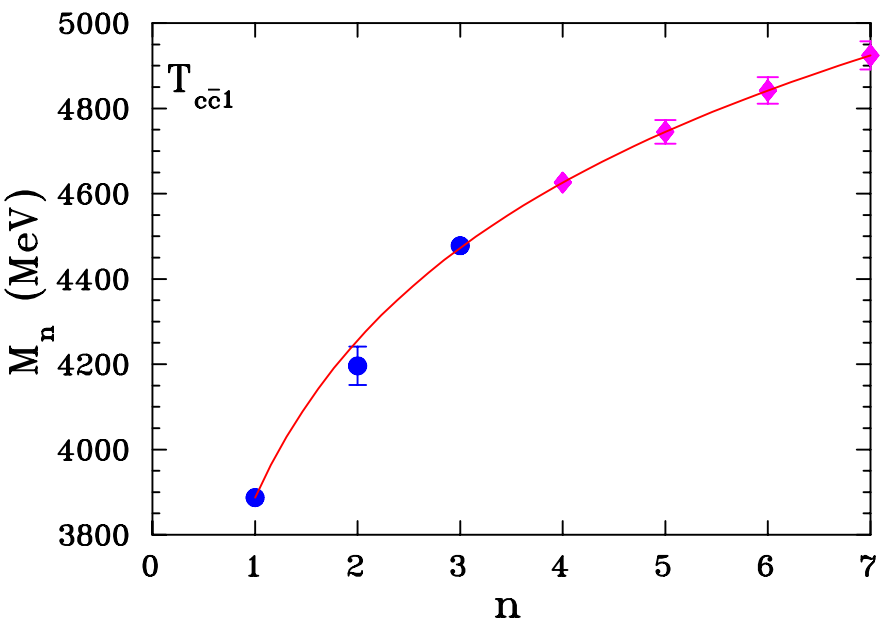} 
}

\centerline{\parbox{0.8\textwidth}{
\caption[] {\protect\small
Data for $T_{c\bar{c}1}$ (blue circles): 
$T_{c\bar{c}1}(3887)$, $T_{c\bar{c}1}(4196)$, and  
$T_{c\bar{c}1}(4478)$~\cite{ParticleDataGroup:2024cfk}. 
Predicted states (magenta diamonds): 
$T_{c\bar{c}1}(4626)$, $T_{c\bar{c}1}(4745)$, $T_{c\bar{c}1}(4842)$, and $T_{c\bar{c}1}(4924)$ with masses of $4626\pm 24~\mathrm{MeV}$, $4745\pm 28~\mathrm{MeV}$, $4842\pm 31~\mathrm{MeV}$, and $4924\pm 33~\mathrm{MeV}$, respectively. The solid red curve presents the best-fit. The fit parameters are $\alpha = 533.0\pm 15.2~\mathrm{MeV}$ and $\beta = 3887.0\pm 
2.6~\mathrm{MeV}$. $\chi^2/\mathrm{DoF} = 1.9$ and CL = 16.7\%.
}}
\label{fig:fig39} } 
\end{figure}

\subsection{$b\bar{b}$ Mesons}

\subsubsection{$\Upsilon(1^{-~-})$ Excited-States}

A series of excited $\Upsilon$-meson states is recorded in the Particle Data Listings~\cite{ParticleDataGroup:2024cfk}. $\Upsilon$: 
$I^G(J^{PC}) = 0^-(1^{-~-})$, $b\bar{b}$.  (Analog to the $q\bar{q}\;\omega$ meson.)

The logarithmic fit to the BW masses (MeV) of the three known $\Upsilon$-meson excited states (blue circles) and four projected higher excited states (magenta diamonds) is shown in 
Fig.~39. PDG gives very small uncertainties for heavy-mass mesons of approximately 0.02\% that do not allow Eq.~(\ref{eq:eq2}) to fit them well. We increased the uncertainties of the input data by a factor of 10 to present the results.
\begin{figure}[htb!]
\centering
{
    \includegraphics[width=0.5\textwidth,keepaspectratio]{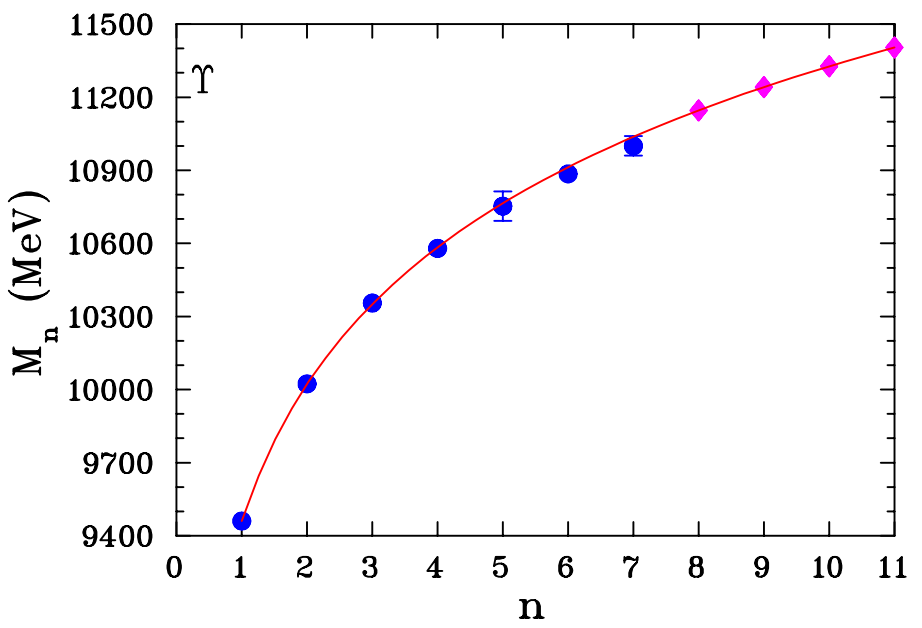} 
}

\centerline{\parbox{0.8\textwidth}{
\caption[] {\protect\small
Data for $\Upsilon(1^{-~-})$ (blue circles): $\Upsilon(9460)$, $\Upsilon(10023)$, $\Upsilon(10355)$, $\Upsilon(10579)$, $\Upsilon(10753)$, $\Upsilon(10885)$, and $\Upsilon(11000)$~\cite{ParticleDataGroup:2024cfk}.
Predicted states (magenta diamonds): $\Upsilon(11146)$, $\Upsilon(11241)$, $\Upsilon(11327)$, and $\Upsilon(11404)$ with masses of $11146\pm 13~\mathrm{MeV}$, $11241\pm 14~\mathrm{MeV}$, $11327\pm 14~\mathrm{MeV}$, and $11404\pm 15~\mathrm{MeV}$, respectively. (Note that the uncertainties are so small that many more states can be predicted to high accuracy.)
The solid red curve presents the best-fit.
The fit parameters are $\alpha = 810.6\pm 3.4~\mathrm{MeV}$ and $\beta = 9460.5\pm 
1.0~\mathrm{MeV}$.
$\chi^2/\mathrm{DoF} = 0.7$ and CL = 61.5\%. Taking into account the fact that we increased uncertainties of input data by a factor of 10.
}
\label{fig:fig40} } }
\end{figure}

\subsubsection{$\chi_b(1)(1^{++})$ Excited-States}

A series of excited $\chi_b(1)$-meson states is recorded in the Particle Data Listings~\cite{ParticleDataGroup:2024cfk}. $\chi_b(1)$: 
$I^G(J^{PC}) = 0^+(1^{++})$, $b\bar{b}$.  (Analog to $q\bar{q}\;f_1$ meson.)

The logarithmic fit to the BW masses (MeV) of the four known $\chi$-meson excited states (blue circles) and four projected higher excited states (magenta diamonds) is shown in Fig.~40. 
PDG gives very small uncertainties for heavy-mass mesons of approximately 0.01\% that do not allow Eq.~(\ref{eq:eq2}) to 
fit them well. We increased the uncertainties of the input data by a factor of 40 to present the results.
\begin{figure}[htb!]
\centering
{
    \includegraphics[width=0.5\textwidth,keepaspectratio]{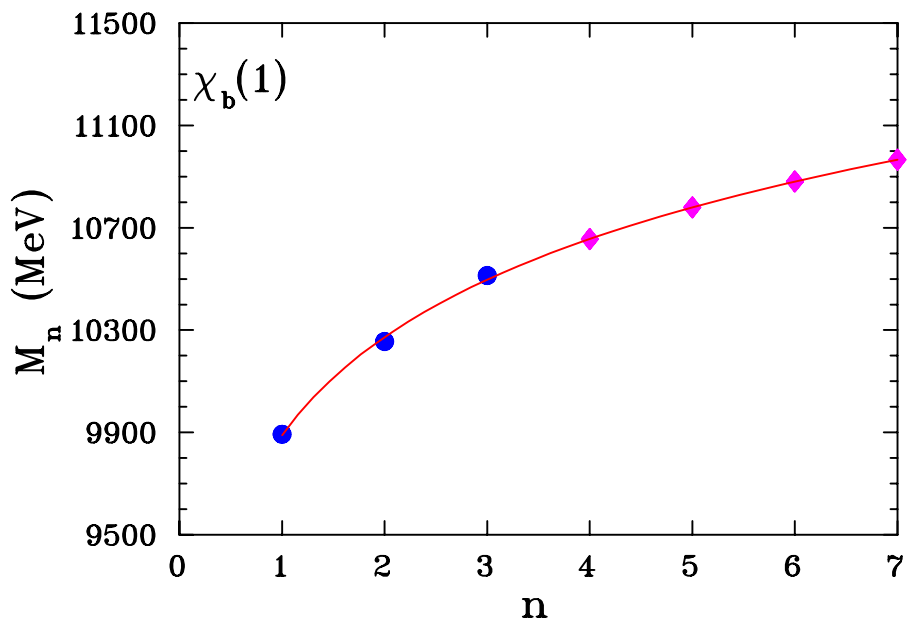} 
}

\centerline{\parbox{0.8\textwidth}{
\caption[] {\protect\small
Data for $\chi_b(1)(1^{++})$ (blue circles): $\chi_b(1)(9893)$, $\chi_b(1)(10255)$, and  
$\chi_b(1)(10513)$~\cite{ParticleDataGroup:2024cfk}. 
Predicted states (magenta diamonds): $\chi_b(1)(10656)$, $\chi_b(1)(10778)$, $\chi_b(1)(10880)$, and $\chi_b(1)(10966)$ with masses of $10656\pm 29~\mathrm{MeV}$, $10780\pm 31~\mathrm{MeV}$, $10881\pm 33~\mathrm{MeV}$, and $10966\pm 34~\mathrm{MeV}$, respectively. (Note that the uncertainties are so small that many more states can be predicted to high accuracy.)
The solid red curve presents the best-fit.
The fit parameters are $\alpha = 553.2\pm 26.0\mathrm{MeV}$ and $\beta = 9889.4\pm 
15.6~\mathrm{MeV}$.
$\chi^2/\mathrm{DoF} = 1.0$ and CL = 30.8\%, taking into account the fact that we increased uncertainties of input data by a factor of 40.
}}
\label{fig:fig41} } 
\end{figure}

\subsubsection{$\chi_b(2)(2^{++})$ Excited-States}

A series of excited $\chi_b(2)$-meson states is recorded in the Particle Data Listings~\cite{ParticleDataGroup:2024cfk}. $\chi_b(2)$: 
$I^G(J^{PC}) = 0^+(2^{++})$, $b\bar{b}$.  (Analog to $q\bar{q}\;f_2$ meson.)

The logarithmic fit to the BW masses (MeV) of the four known $\chi$-meson excited states (blue circles) and four projected higher excited states (magenta diamonds) is shown in Fig.~41. 
PDG gives very small uncertainties for heavy-mass mesons of approximately 0.01\% that do not allow Eq.~(\ref{eq:eq2}) to 
fit them well. We increased the uncertainties of the input data by a factor of 40.
\begin{figure}[htb!]
\centering
{
    \includegraphics[width=0.5\textwidth,keepaspectratio]{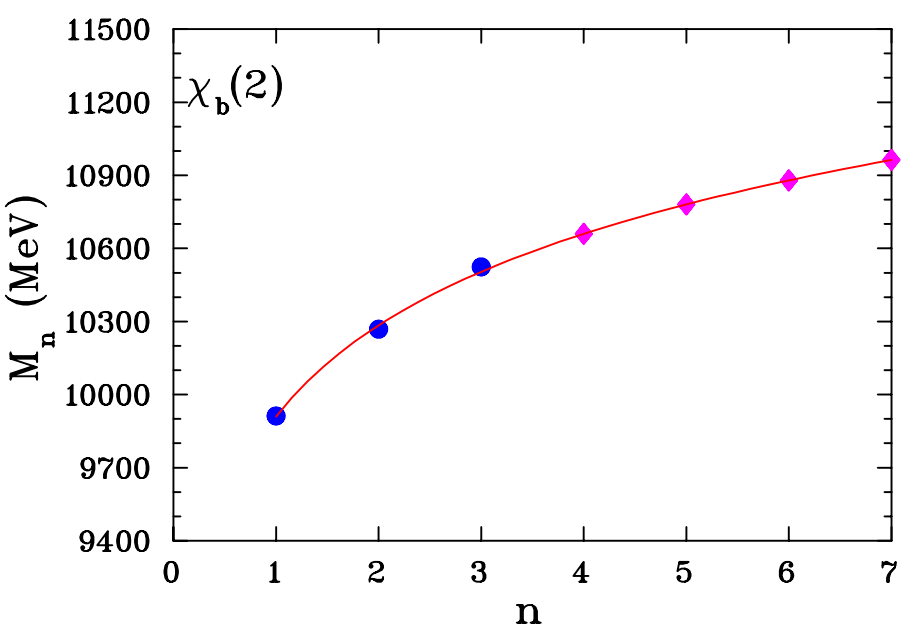} 
}

\centerline{\parbox{0.8\textwidth}{
\caption[] {\protect\small
Data for $\chi_b(2)(2^{++})$ (blue circles): $\chi_b(2)(9912)$, $\chi_b(2)(10269)$, and  
$\chi_b(2)(10524)$~\cite{ParticleDataGroup:2024cfk}. 
Predicted states (magenta diamonds): $\chi_b(2)(10660)$, $\chi_b(2)(10781)$, $\chi_b(2)(10879)$, and $\chi_b(2)(10963)$ with masses of $10660\pm 32~\mathrm{MeV}$, $10781\pm 35~\mathrm{MeV}$, $10879\pm 37~\mathrm{MeV}$, and $10963\pm 39~\mathrm{MeV}$, respectively. (Note that the uncertainties are so small that many more states can be predicted to high accuracy.)
The solid red curve presents the best-fit.
The fit parameters are $\alpha = 541.5\pm 37.8~\mathrm{MeV}$ and $\beta = 9909.1\pm 
15.7~\mathrm{MeV}$.
$\chi^2/\mathrm{DoF} = 1.0$ and CL = 32.6\%, taking into account the fact that we increased uncertainties of input data by a factor of 40.
}}
\label{fig:fig42} } 
\end{figure}


\subsection{Cumulative Meson Excited-States}

The cumulative fit curves of twenty-four meson equal-quantum mass sets are shown in Fig.~42.
\begin{figure}[htb!]
\centering
{
    \includegraphics[width=0.44\textwidth,keepaspectratio]{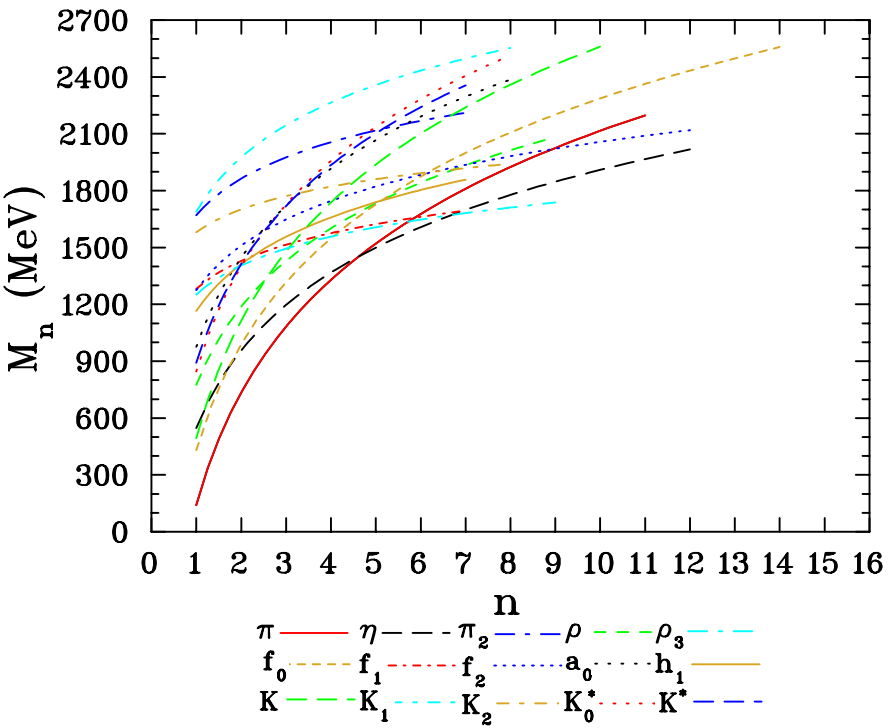}~~~ 
    \includegraphics[width=0.47\textwidth,keepaspectratio]{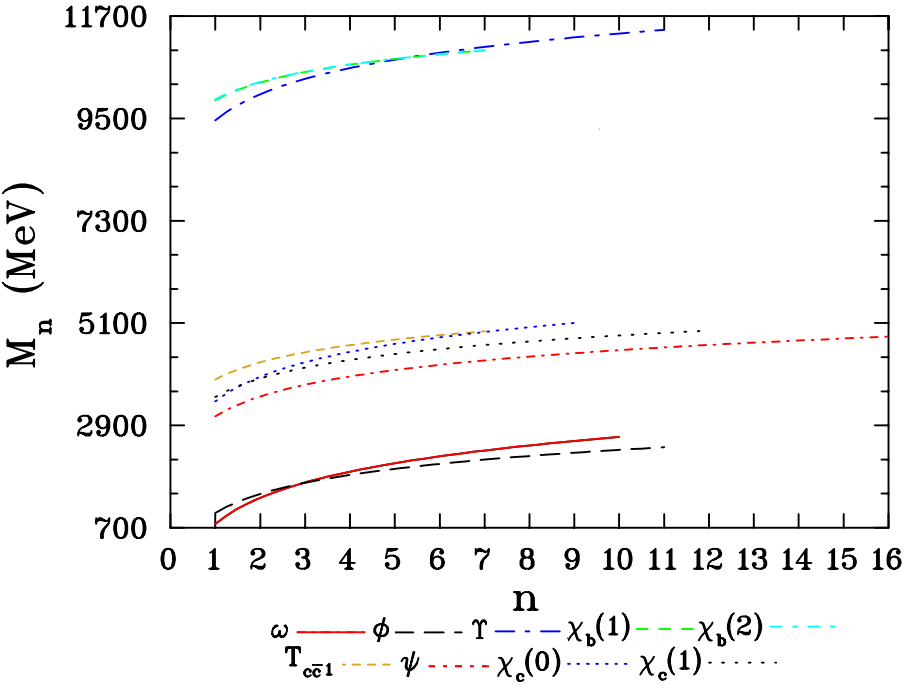}
}

\centerline{\parbox{0.8\textwidth}{
\caption[] {\protect\small
Cumulative fit curves for the excited states of the meson. 
\underline{Left}: $\pi(0^{-+})$, $\eta(0^{-+})$, $\pi_2(2^{-+})$, 
$\rho(1^{-~-})$, $\rho_3(3^{-~-})$, $f_0(0^{++})$, $f_1(1^{++})$, 
$f_2$(2$^{++}$), $a_0(0^{++})$, $h_1(1^{+-})$, $K(0^-)$,
$K_1(1^+)$, $K_2(2^-)$, $K^\ast_0(0^ +)$, and $K^\ast(1^-)$.
\underline{Right}: $\omega(1^{-~-})$, $\phi(1^{-~-})$, 
$\psi(1^{-~-})$, $\chi_{c0}(0^{++})$, $\chi_{c1}(1^{++})$, 
$\Upsilon(1^{-~-})$, $\chi_{b1}(1^{++})$, 
$\chi_{b2}(2^{++})$, and $T_{c\bar{c}1}$.
The two curves  $\chi_{b1}(1^{++})$ and 
$\chi_{b2}(2^{++})$ are very nearly the same and are not visually separable in the graph.
Note that the ``logarithmic slope,'' $\alpha$ in Eq.~(\ref{eq:eq2}), usually decreases as the ground-state mass increases.
}}
\label{fig:mes} } 
\end{figure}
\section{Potential-Energy Function}
\begin{figure}
    \centering
        \includegraphics[width=0.45\linewidth]{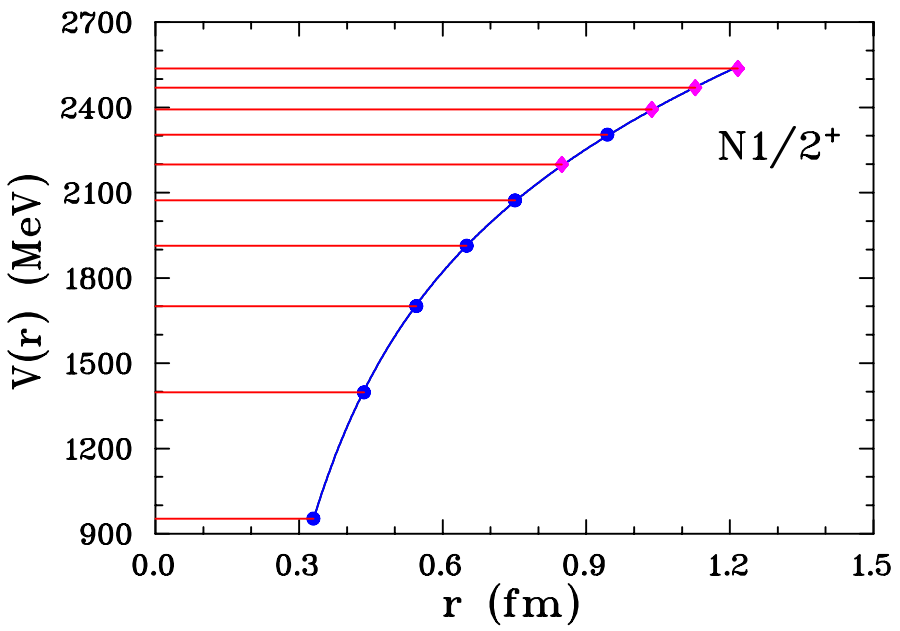}~~
        \includegraphics[width=0.45\linewidth]{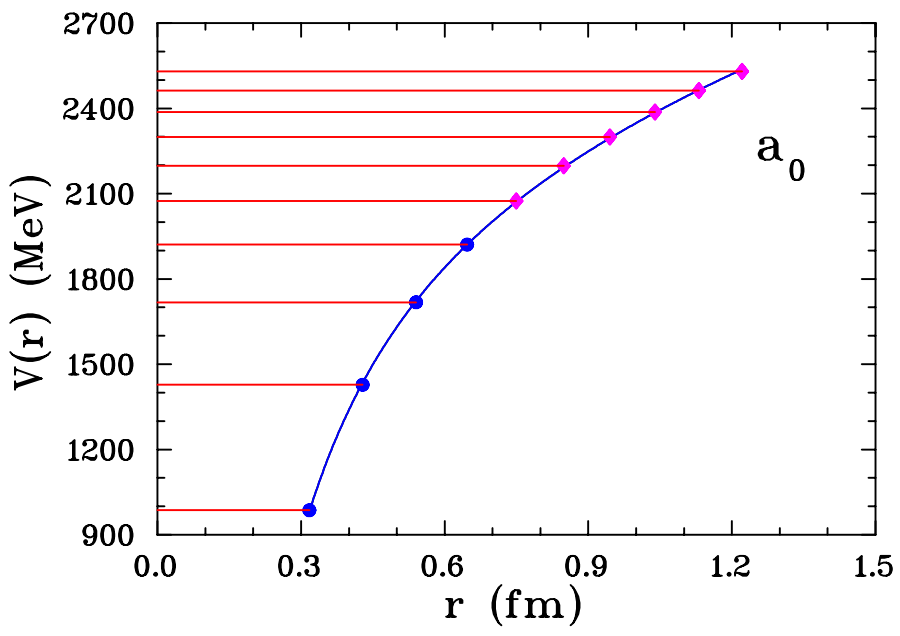}
    \caption{$N1/2^+$ baryon excited-states set (left) and $a_0$  meson excited-states set (right) fits using the Cornell potential-energy function, Eq.~(\ref{eq:corn}), with radii created by using increasing number of de Broglie wavelengths by half wavelengths with increasing mass levels. The fits are generated using energy levels up to $n = 10$. The parameter values for the two states sets are 
    $N1/2^+$: $A = 418.26~\mathrm{MeV\cdot fm}$, 
    $B = 407.52~\mathrm{MeV/fm}$, 
    and $C = 2506.2~\mathrm{MeV}$;     
    $a_0$: $A = 369.23~\mathrm{MeV\cdot fm}$, 
    $B = 450.27~\mathrm{MeV/fm}$, 
    and $C = 2390.5~\mathrm{MeV}$. PDG data~\cite{ParticleDataGroup:2024cfk} by blue circles. Magenta diamonds are predicted states.}
    \label{fig:N12Pot}
\end{figure}

The $N1/2^+$ baryon (Fig.~\ref{fig:N12Pot} (left)) and the $a_0$ meson (Fig.~\ref{fig:N12Pot} (right)) are used to illustrate how their energy spectra can be used to approximate the shape of their potential-energy functions.

The approximate quantum principles used here are the de Broglie wavelength (DBWL) for each mass/energy level, $\lambda_n = \frac{hc}{M_n} = 2\pi\frac{\hbar c}{M_n}$, where $M_n$ is in the energy unit [MeV] and $\hbar c = 197.327~\mathrm{MeV\cdot fm}$, and the assumption that the mass levels' diameters occupy integers times half the DBWLs across the circles of the spherical radial potential energy (PE). The approximate potential-energy plot versus radius is $M_n$ \textit{vs.} $r_n$ where, in [fm],
\begin{equation}
   r_n=n\frac{\lambda_n}{4} = n\frac{\pi}{2}\frac{\hbar c}{M_n} \>,
\label{eq:radius}
\end{equation}
where $n$ is the radial quantum number and the number of half-wavelengths in a diameter.
This potential-energy curve is similar to the Cornell-potential function~\cite{Eichten:1974af, Eichten:1978tg, Brambilla:1999ja} used in lattice QCD calculations~\cite{Yamamoto:2008jz}: 
\begin{equation}
    V(r) = -\frac{4}{3}A / r + B~r + C \>,
\label{eq:corn}
\end{equation}
where $r$ is the effective radius of the resonance state and $A$, $B$, and $C$ are free parameters. The infinitely increasing PE curve provides confinement for constituents.

Two fits for $V(r)$ are shown in Fig.~\ref{fig:N12Pot} for the sets of $N1/2^+$ baryon and $a_0$ meson excited states. These fits have
Eq.~(\ref{eq:eq1}) equal to Eq.~(\ref{eq:corn}) at high accuracy for masses starting at $M_1$ up to the highest mass at which the fit is made, in the two cases here $M_{10}$; however, mass values below $M_1$ do not exist according to Eq.~(\ref{eq:eq1}) because $Ln(n)$ for the integer $n<1$ does not exist. The $-\frac{1}{r}$ first term in Eq.~(\ref{eq:corn}) will cause entry into negative numbers at a radius considerably below the lowest fit radius.

Calculation of $b\bar{b}$ equal-quantum excited-states sets' masses have been performed by Kher \textit{et al.}~\cite{Kher:2022cp} using the Cornell potential and Gaussian wavefunctions; the results do not fit the data as well as our logarithmic universal mass equation.

These potential-energy (PE) curves are approximations because the wave function  is not simply a train of half de Broglie wavelengths. The calculation yields the general shape of the PE, as the authors have shown by calculating for the simple Bohr hydrogen atom energy levels (yields $\approx-\frac{1}{r}$ shape) and the 3D-oscillator energy levels (yields $\approx$ 3D-parabola shape).

\section{Conclusion}
\label{Sec:Conc}

Summary Tables~\ref{tbl:taba1} and \ref{tbl:taba2} for free parameters $\alpha$ and $\beta$ of Eq.~(\ref{eq:eq2}) represented the best-fit results for the baryon and meson sets of data show that the parameter $\beta$ corresponds to the mass of the ground resonance.

Some interesting results of this study are:
\begin{enumerate}
\item Logarithmic behavior of the masses of particles with the same quantum numbers at fixed $J^P$ for baryons and fixed $J^{PC}$ for mesons,
\item Prediction of several missing states~\cite{Koniuk:1979vw},
\item Prediction of four higher-mass~excited states for each of the 39 data sets; \textit{i.e.}, $39\times 4 = 156$ higher-mass excited-states are predicted. In addition, our fits allow us to determine the lesser masses of 53
states missing in the PDG. We found that the masses of a small number of states reported by PDG do not fit our simple Eq.~(\ref{eq:eq2}) and we omitted them in the fits.
\item Cornell potential is an example of how a logarithmic behavior can be explained by an appropriate potential.
\end{enumerate}

There are many excited states reported in the Particle Data Listings that have less than three states in a set of equal quantum numbers, or that have three or more states in a set with masses that do not behave according to a logarithmic function.

The authors conclude that the 39 excited states equal-quantum mass data sets that fit so well to the logarithm function, as shown in this document, imply that the logarithm is a general function for all excited states equal-quantum mass data sets (including the $s\bar{s}$, $c\bar{c}$ and $b\bar{b}$ states); at least for the mass range of currently known excited states. Thus, a universal mass equation for equal-quantum excited states sets is presented.

After the two lowest-mass excited states in equal-quantum excited-states sets are reported at high accuracy from experiments, our universal mass equation for equal-quantum excited-states sets can be used to predict several higher-mass states in the set.

Fixing the free parameter $\beta$ in Eq.~(\ref{eq:eq2}) as the mass of the ground state,  two accurate masses that start a set can be used to accurately calculate four higher masses in the set~\cite{Roper:2025}.

The proposed two-parameter logarithm function is an opportunity to look for missed baryon- and meson-resonances predicted by QCD models and LQCD calculations. That is one of the goals of modern experiments at LHC, BESIII, JLab, and J-PARC.

\begin{table}[htb!]

\centering \protect\caption{{Summary for free parameters $\alpha$ and $\beta$ of Eq.~(\ref{eq:eq2}) corresponding to the best-fit results for the 15 excited baryon states' sets. The PDG's BW mass 
estimation is within uncertainty of the obtained parameter $\beta$. Question marks indicate that the PDG does not report $P^J$$^\ddagger$.}}

\vspace{2mm}
{%
\begin{tabular}{|c|c|c|c|c|}
\hline
Baryon          & Ground-State Mass &  $\alpha$      &  $\beta$  & CL \tabularnewline
                &   (MeV)           & (MeV)          & (MeV)     & (\%) \tabularnewline
\hline
$P_{c\bar{c}}^+?^?$$^\ddagger$&4312 &  90.7$\pm$2.9  & 4312.2$\pm$3.8     & 82.6 \tabularnewline
$N1/2^-$        &   1530            & 200.7$\pm$16.0 & 1522.7$\pm$13.3    & 29.1 \tabularnewline
$\Lambda 3/2^-$ &   1519            & 250.4$\pm$6.1  & 1518.9$\pm$1.0     & 36.9 \tabularnewline
$N5/2^+$        &   1685            & 287.7$\pm$27.1 & 1685.0$\pm$5.0     & 96.3 \tabularnewline
$\Sigma 1/2^-$  &   1620            & 289.1$\pm$27.1 & 1610.4$\pm$24.0    & 34.8 \tabularnewline
$N3/2^+$        &   1720            & 304.4$\pm$39.1 & 1716.3$\pm$32.7    & 76.7 \tabularnewline
$N3/2^-$        &   1515            & 332.0$\pm$20.9 & 1514.9$\pm$5.0     & 91.8 \tabularnewline
$\Delta 1/2^-$  &   1610            & 373.0$\pm$50.6 & 1609.0$\pm$19.8    & 75.8 \tabularnewline
$\Lambda 1/2^-$ &   1405            & 387.5$\pm$6.0  & 1405.1$\pm$1.3     & 53.5 \tabularnewline
$\Lambda 1/2^+$ &   1116            & 430.9$\pm$8.5  & 1115.700$\pm$0.006 & 86.9 \tabularnewline
$\Delta 3/2^+$  &   1232            & 495.3$\pm$34.0 & 1232.0$\pm$2.0     & 93.6 \tabularnewline
$N5/2^-$        &   1675            & 501.0$\pm$15.6 & 1675.4$\pm$8.0     & 35.8 \tabularnewline
$\Sigma 3/2^+$  &   1383            & 519.4$\pm$10.6 & 1382.8$\pm$0.3     & 61.4 \tabularnewline
$\Sigma 1/2^+$  &   1189            & 668.0$\pm$25.5 & 1189.40$\pm$0.07   & 41.5 \tabularnewline
$N1/2^+$        &    940            & 698.2$\pm$14.1 & 939.6$\pm$5.4$\times 10^{-7}$ & 68.7 \tabularnewline
\hline
\end{tabular}} \label{tbl:taba1}
\end{table}

\begin{table}[htb!]

\centering \protect\caption{{Summary for free parameters $\alpha$ and $\beta$ of Eq.~(\ref{eq:eq2}) 
corresponding to the best-fit results for the 24 excited meson states sets. The PDG's BW mass 
estimation is within uncertainty of the obtained parameter $\beta$. While the estimated mass for the 
$f_0(600)$ is really broad, 400 to $800~\mathrm{MeV}$~\cite{ParticleDataGroup:2024cfk}. Our best-fit result 
is close to its low limit$^\ddagger$.}
}

\vspace{2mm}
{%
\begin{tabular}{|c|c|c|c|c|}
\hline
Meson            & Ground-State Mass &  $\alpha$      &  $\beta$          & CL \tabularnewline
                 &   (MeV)           & (MeV)          & (MeV)             & (\%)  \tabularnewline
\hline
$K_2(2^-)$       &   1580            & 173.9$\pm$9.9  & 1580.5$\pm$9.9    & 77.5 \tabularnewline
$f_1(1^{++})$    &   1282            & 210.3$\pm$2.0  & 1281.3$\pm$0.5    & 80.8 \tabularnewline
$K_1(1^+)$       &   1253            & 220.8$\pm$13.4 & 1251.9$\pm$6.9    & 36.1 \tabularnewline
$\pi_2(2^{-+})$  &   1671            & 277.8$\pm$15.3 & 1670.7$\pm$2.1    & 37.2 \tabularnewline
$f_2(2^{++})$    &   1275            & 329.6$\pm$2.9  & 1275.3$\pm$0.8    & 18.3 \tabularnewline
$h_1(1^{+-})$    &   1166            & 356.1$\pm$14.7 & 1165.5$\pm$6.0    & 29.3 \tabularnewline
$\rho_3(3^{--})$ &   1689            & 415.5$\pm$16.1 & 1688.9$\pm$2.1    & 55.1 \tabularnewline
$T_{c\bar{c}1}(1^{+-})$& 3887        & 533.0$\pm$15.2 & 3887.0$\pm$2.6    & 16.7 \tabularnewline
$\chi_{b2}(2^{++})$& 9912            & 540.3$\pm$19.4 & 9908.8$\pm$11.2   & 14.8 \tabularnewline
$\chi_{b1}(1^{++})$& 9893            & 553.2$\pm$26.0 & 9889.4$\pm$15.6   & 30.8 \tabularnewline
$\chi_{c1}(1^{++})$& 3511            & 574.8$\pm$4.8  & 3510.7$\pm$0.1    & 30.6 \tabularnewline
$\phi(1^{-~-})$  &   1019            & 588.5$\pm$3.0  & 1019.50$\pm$0.02  & 47.9 \tabularnewline
$\eta(0^{-+})$   &    548            & 591.39$\pm$0.09&  547.86$\pm$0.02  & 46.6 \tabularnewline
$\rho(1^{-~-})$  &    775            & 595.4$\pm$10.47&  775.3$\pm$0.2    & 26.3 \tabularnewline
$\psi(1^{-~-})$  &   3097            & 617.3$\pm$2.7  & 3096.90$\pm$0.03  & 25.4 \tabularnewline
$a_0(0^{++})$    &    980            & 677.8$\pm$20.6 &  976.8$\pm$19.2   & 74.6 \tabularnewline
$K^\ast(1^-)$    &    892            & 752.7$\pm$13.1 &  891.7$\pm$0.3    & 95.9 \tabularnewline
$\chi_{c0}(0^{++})$& 3415            & 768.5$\pm$13.3 & 3414.7$\pm$1.5    & 66.7 \tabularnewline
$K^\ast_0(0^+)$  &    845            & 801.8$\pm$15.9 &  846.3$\pm$16.8   & 63.9  \tabularnewline
$f_0(0^{++})$    &    600            & 805.8$\pm$13.6 &  431.3$\pm$24.2$^\ddagger$  & 22.7 \tabularnewline
$\Upsilon(1^{-~-})$& 9460            & 810.6$\pm$3.4  & 9460.5$\pm$1.0    & 61.5 \tabularnewline
$\omega(1^{-~-})$&    783            & 811.3$\pm$15.0 &  782.7$\pm$0.1    & 54.7 \tabularnewline
$\pi(0^-)$       &    139            & 858.3$\pm$5.1  &  139.6$\pm$1.8$\times 10^{-4}$ & 76.8 \tabularnewline
$K(0^-)$         &    494            & 897.3$\pm$14.2 &  493.68$\pm$0.02& 49.8 \tabularnewline
\hline
\end{tabular}} \label{tbl:taba2}
\end{table}

Unfortunately, we have no explanation for this phenomenon. The problem is that within the potential model, one has to build a potential which {\em simultaneously} describes the $n$-dependence of radial excitation masses and the $J$-dependence of resonances on the Regge trajectory. On the other hand, the logarithmic n-dependence, presented here, is an interesting observation which should be accounted for in any future theory of hadronic states.


\section*{Acknowledgments}

This work was supported in part by the U.~S.~Department of Energy,  Office of Science, Office of Nuclear Physics, under Award No.~DE--SC0016583.
L.~David~Roper wishes to express a deep debt of gratitude to Igor~Strakovsky for using his knowledge and many hours of effort to make a rough document into a much better document in particle-physics phenomenology.

Both authors dedicate this document to the memory of their excellent colleague and friend Richard Allen Arndt.



\begin{thebibliography}{99}
\bibitem{ParticleDataGroup:2024cfk}
    S.~Navas \textit{et al.} [Particle Data Group],
    ``Review of particle physics,''
    Phys.\ Rev.\ D\ \textbf{110}, 030001 (2024).
\bibitem{Amsler:2024}
    C.~Amsler \textit{et al} ``8.~Naming Scheme for Hadrons,'' in:
    S.~Navas \textit{et al.} [Particle Data Group],
    ``Review of particle physics,''
    Phys.\ Rev.\ D\ \textbf{110}, 030001 (2024).
\bibitem{Roper:1964zza}
    L.~D.~Roper,
    ``Evidence for a $P_{11}$ pion-nucleon resonance at 556~MeV,''
    Phys.\ Rev.\ Lett.\ \textbf{12}, 340 (1964).
\bibitem{Koniuk:1979vw}
    R.~Koniuk and N.~Isgur,
    ``Where have all the resonances gone? An analysis of baryon couplings in a Quark model with Chromodynamics,''
    Phys.\ Rev.\ Lett.\ \textbf{44}, 845 (1980).
\bibitem{LHCb:2019kea}
    R.~Aaij \textit{et al.} [LHCb Collaboration],
    ``Observation of a narrow pentaquark state, $P_c(4312)^+$, and of two-peak structure of the $P_c(4450)^+$,''
    Phys.\ Rev.\ Lett.\ \textbf{122}, 222001 (2019).
\bibitem{LHCb:2015yax}
    R.~Aaij \textit{et al.} [LHCb Collaboration],
    ``Observation of $J/\psi p$ resonances consistent with pentaquark states in $\Lambda_b^0 \to J/\psi K^- p$ decays,''
    Phys.\ Rev.\ Lett.\ \textbf{115}, 072001 (2015).
\bibitem{Eichten:1974af}
    E.~Eichten, K.~Gottfried, T.~Kinoshita, J.~B.~Kogut, K.~D.~Lane, and T.~M.~Yan,
    ``The spectrum of Charmonium,''
    Phys.\ Rev.\ Lett.\ \textbf{34}, 369 (1975) 
    [erratum: Phys.\ Rev.\ Lett.\ \textbf{36}, 1276 (1976)].
\bibitem{Eichten:1978tg}
    E.~Eichten, K.~Gottfried, T.~Kinoshita, K.~D.~Lane, and T.~M.~Yan,
    ``Charmonium: The model,''
    Phys.\ Rev.\ D\ \textbf{17}, 3090 (1978)
    [erratum: Phys.\ Rev.\ D\ \textbf{21}, 313 (1980)].
\bibitem{Brambilla:1999ja}
    N.~Brambilla and A.~Vairo,
    ``Quark confinement and the hadron spectrum,''
    [arXiv:hep-ph/9904330 [hep-ph]].
\bibitem{Yamamoto:2008jz}
    A.~Yamamoto, H.~Suganuma, and H.~Iida,
    ``Lattice QCD study of the heavy-heavy-light quark potential,''
    Phys.\ Rev.\ D\ \textbf{78}, 014513 (2008).
    \bibitem{Kher:2022cp}
    V.~Kher, R.~Chaturvedi, N.~Devlani, and A.~Rai,
    ``Bottomonium spectroscopy using Coulomb plus linear (Cornell) potential,''
    arXiv:2201.08317 [hep-ph].
\bibitem{Roper:2025}
    L.~D.~Roper and I.~Strakovsky,
    ``Universal mass equation for equal-quantum excited-states sets~II,''
    in progress.
\end{thebibliography}
\end{document}